\documentclass[twocolumn,amssymb, nobibnotes, aps, prd,amsmath,mathtools, nofootinbib,pgfmath,pgffor,superscriptaddress]{revtex4-1}
\usepackage{graphicx}

\usepackage{caption}
\usepackage{subcaption}

\usepackage{hyperref}
\hypersetup{colorlinks,citecolor=blue}

\makeatletter
\def\qrr@split@result#1 #2\@qrr@split@result{\edef\erfInput{#1}\edef\erfResult{#2}}
\newcommand*{\gnuplotErf}[2][\jobname.eval]{%
    \immediate\write18{gnuplot -e "set print '#1'; print #2, erf(#2);"}%
    \everyeof{\noexpand}
    \edef\qrr@temp{\input{#1} }%
    \expandafter\qrr@split@result\qrr@temp\@qrr@split@result
}
\makeatother

\usepackage{caption,color}
\usepackage{subcaption}
\usepackage{multirow}

\usepackage{hyperref}
\hypersetup{colorlinks,citecolor=blue}

\begin{document}

\title[Echoes from the Abyss]{Echoes from the Abyss: Tentative evidence for Planck-scale structure at black hole horizons}

\author{Jahed Abedi}
\email{jahed$_$abedi@physics.sharif.ir}
\affiliation{Department of Physics, Sharif University of Technology, P.O. Box 11155-9161, Tehran, Iran}
\affiliation{School of Particles and Accelerators, Institute for Research in Fundamental Sciences (IPM), P.O. Box 19395-5531, Tehran, Iran}
\affiliation{Perimeter Institute for Theoretical Physics, 31 Caroline St. N., Waterloo, ON, N2L 2Y5, Canada}

\author{Hannah Dykaar}
\affiliation{Department of Physics, McGill University, 3600 rue University, Montreal, QC, H3A 2T8, Canada}
\affiliation{Department of Physics and Astronomy, University of Waterloo, Waterloo, ON, N2L 3G1, Canada}

\author{Niayesh Afshordi}
\email{nafshordi@pitp.ca}
\affiliation{Perimeter Institute for Theoretical Physics, 31 Caroline St. N., Waterloo, ON, N2L 2Y5, Canada}
\affiliation{Department of Physics and Astronomy, University of Waterloo, Waterloo, ON, N2L 3G1, Canada}

\begin{abstract}
In classical General Relativity (GR), an observer falling into an astrophysical black hole is not expected to experience anything dramatic as she crosses the event horizon. However, tentative resolutions to problems in quantum gravity, such as the cosmological constant problem, or the black hole information paradox, invoke significant departures from classicality in the vicinity of the horizon. It was recently pointed out that such near-horizon structures can lead to late-time echoes in the black hole merger gravitational wave signals that are otherwise indistinguishable from GR. We search for observational signatures of these echoes in the gravitational wave data released by advanced Laser Interferometer Gravitational-Wave  Observatory (LIGO), following the three black hole merger events GW150914, GW151226, and LVT151012. In particular, we look for repeating damped echoes with time-delays of $8 M \log M$ (+spin corrections, in Planck units), corresponding to Planck-scale departures from GR near their respective horizons. Accounting for the ``look elsewhere'' effect due to uncertainty in the echo template, we find tentative evidence for Planck-scale structure near black hole horizons at false detection probability of $1\%$ (corresponding to $2.5\sigma$\footnote{In this {\it paper}, we use 2-tailed gaussian probability to assign a significance to a p-value, e.g., $1-$p-value$=$ 68\% and 95\% correspond to 1$\sigma$ and 2$\sigma$ respectively.} significance level). 
Future observations from interferometric detectors at higher sensitivity, along with more physical echo templates, will be able to confirm (or rule out) this finding, providing possible empirical evidence for alternatives to classical black holes, such as in {\it firewall} or {\it fuzzball} paradigms.
\end{abstract}

\maketitle

\section{\label{Introduction}Introduction}

There is mounting, albeit controversial,  theoretical evidence that  quantum black holes might be significantly different from their classical counterparts, even in the regime where semi-classical gravity is expected to be valid. Such strong modifications may exist, not only due to non-perturbative quantum gravitational effects \cite{Almheiri:2012rt,Lunin:2001jy,Lunin:2002qf,Maldacena:2013xja}, but also at the level of semi-classical approximation \cite{Abedi:2013xua,Abedi:2015yga}. In particular, modern versions of Hawking's black hole information paradox have led to exotic alternatives to classical  black hole horizons such as  fuzzball \cite{Lunin:2001jy,Lunin:2002qf} and firewall paradigms  \cite{Braunstein:2009my,Almheiri:2012rt}. These should form by Page time $\sim M^3$, but may emerge as early as the ``scrambling time'' $\sim M \log M$ \cite{Hayden:2007cs, Sekino:2008he}, where $M$ is the black hole mass in Planck units.

On more phenomenological grounds, it has been proposed that a wholesome solution to the (old and new) cosmological constant problems replaces the black hole horizons by a  Planck-scale quantum barrier, which could naturally explain the observed scale of dark energy \cite{PrescodWeinstein:2009mp}. Furthermore, accretion into these ``black holes'' offers a possible origin for observed ultra high energy IceCube neutrinos \cite{Afshordi:2015foa}. 

In this {\it paper}, we search for possible signatures of quantum gravitational alternatives to black hole horizons in the gravitational wave data releases of black hole mergers observed by the advanced Laser Interferometer Gravitational-Wave Observatory (LIGO) \cite{TheLIGOScientific:2016pea,TheLIGOScientific:2016agk,Giddings:2016tla}. As a simple toy model, we replace the event horizon by a mirror (with Dirichlet boundary conditions) at $\sim$ Planck proper length outside the horizon.  This picture is motivated by the realization that a thermal membrane on the stretched horizon,  satisfying Israel junction conditions with $Z_2$ symmetry, happens to have  a thermal entropy equal to the Bekenstein-Hawking area law \cite{Saravani:2012is}. Therefore, any horizonless microscopic model of the black hole which accounts for its entropy, should act as a mirror, at least for linear long wavelength perturbations. The mirror is not perfect for particles with $\omega \gg T_H$ (= Hawking temperature), as they can excite the microstates of the system, and thus be absorbed by the membrane \cite{Mathur:2012jk}, but should be reflective at $\omega \lesssim T_H$ as these microstates cannot be excited. Incidentally, this is the frequency regime for gravitational waves in the ringdown phase of black hole mergers. In contrast, electromagnetic emissions from accretion into black holes are at much higher frequencies, where the membrane is expected to be highly absorbing, consistent with astrophysical observations \cite{Broderick:2009ph,Broderick:2015tda} (but also see  \cite{Afshordi:2015foa,Pen:2013qva}).  
\begin{figure}
\includegraphics[width=0.5\textwidth]{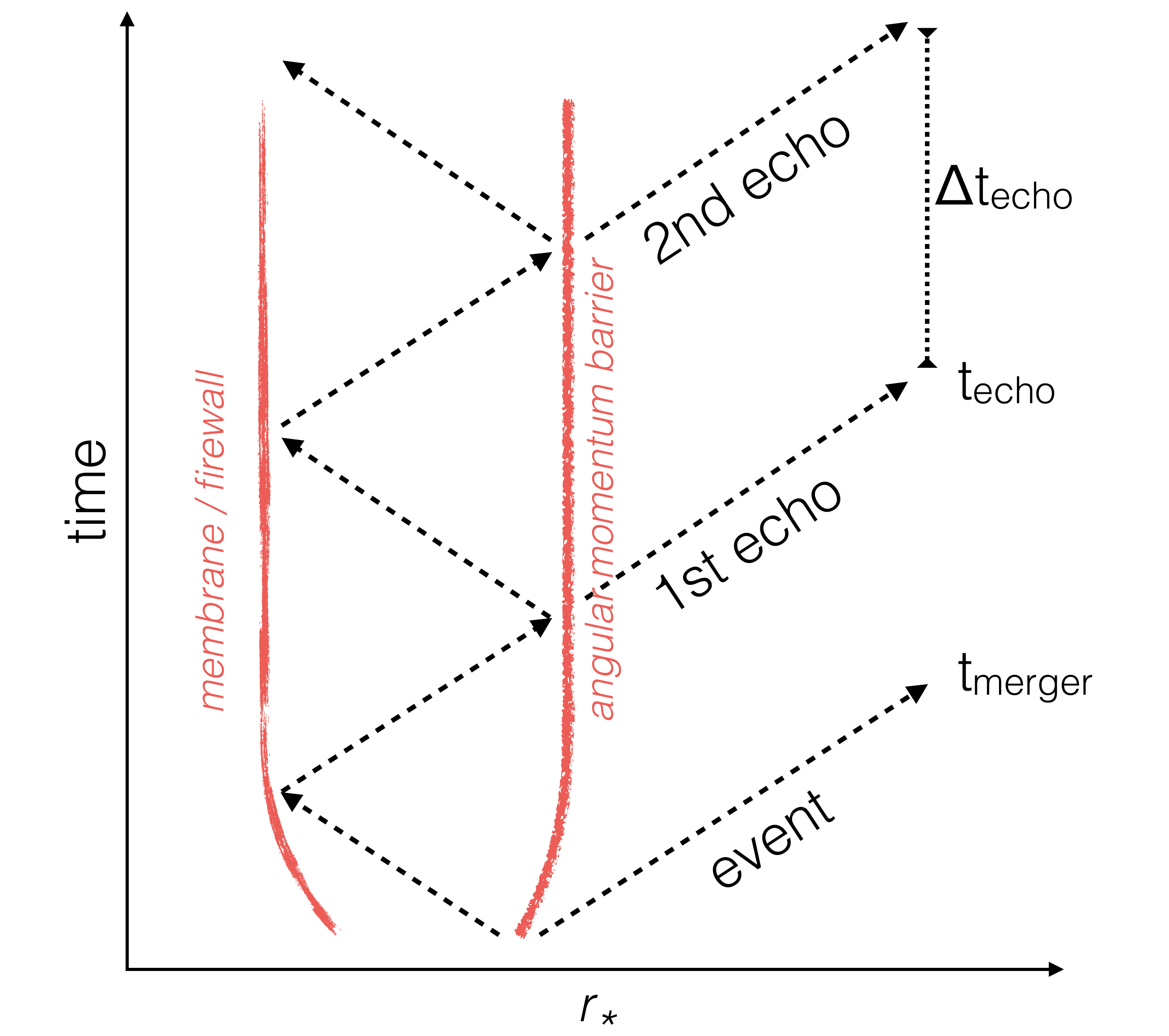}
\caption{Spacetime depiction of gravitational wave echoes from a membrane/firewall on the stretched horizon, following a black hole merger event. }
\label{echo_pic_1}
\end{figure}

In spite of its simplicity, this picture is remarkably robust: As first noticed in \cite{Cardoso:2016rao,Cardoso:2016oxy}, introduction of structure near event horizon leads to late, repeating, echoes of the ringdown phase of the black hole merger, due to waves trapped between the near-horizon structure and the angular momentum barrier (Fig. \ref{echo_pic_1}).   This is relatively insensitive to the nature of the structure, or indeed how one defines the Planck length, $l_p$, as the time for reflection from the stretched horizon is only logarithmically dependent on its distance from the event horizon, i.e. $\Delta t_{\rm echo}=8M\log(M/l_p)$ (+ spin corrections; see below). As a result, e.g., an order of magnitude change in this distance only affects the time of the echoes at $2-3\%$ level. While $\Delta t_{\rm echo}$ is determined by linear physics, the time between the main merger event and the first echo could be further affected by non-linear physics during merger, i.e. $t_{\rm echo}-t_{\rm merger}=\Delta t_{\rm echo}+{\cal O}(M)$ (see Fig. \ref{echo_pic_1}), or equivalently:
\begin{equation}
\frac{t_{\rm echo}-t_{\rm merger}}{\Delta t_{\rm echo}}=1\pm {\cal O}(1 \%),
\label{nonlinear}
\end{equation}
where $\Delta t_{\rm{echo}}$ is predicted from the final (redshifted) mass and spin measurements for each event.

Quite surprisingly, we find statistical evidence for these delayed echoes in LIGO events: GW150914, GW151226, and LVT151012 at a false detection probability of $1\%$ or combined significance of $2.5\sigma$. We shall first  describe our theoretical framework for the echoes, and then our statistical methodology and results.

\section{Echo time-delays} 

At the linear order, perturbed black holes are described by quasi-normal modes (QNM's) which satisfy the boundary conditions of purely outgoing waves at infinity and purely ingoing waves at the horizon. The transition (from ingoing to outgoing) takes place continuously at the peak of the black hole angular momentum potential barrier.

In our case, the ingoing modes of the ringdown reflect back from the membrane (e.g., fuzzball or firewall) near horizon and pass back through the potential barrier. Part of the wave goes to infinity with a time delay. We call this the 1st echo (see Fig. \ref{echo_pic_1}). This time delay corresponds to twice the tortoise coordinate distance between the peak of the angular momentum  barrier ($r_{\rm max}$) and the membrane (which diverges logarithmically if the membrane approaches the horizon) . The remaining part of the 1st echo returns back towards the membrane and the process repeats itself \footnote{Also, note that due to the different boundary conditions near the horizon (compared to the classical picture) there exist a completely different QNM spectrum. A coherent superposition of a large number of these modes is responsible for creating echoes \cite{Cardoso:2016rao,Price:2017cjr,Kokkotas1999}}.  
Assuming Dirichlet boundary conditions at the membrane (discussed above), the reflected waves must be phase inverted, i.e. even echoes have opposite phase with respect to the odd ones (a similar phase flip pattern is also observed in \cite{Cardoso:2016oxy}).

For Kerr black hole with dimensionless spin parameter $a$, this implies:
\begin{eqnarray}
\Delta t_{\rm echo}
=2 \times r_{*}|_{r_{+}+\Delta r}^{r_{\rm max}}=2\times \int_{r_{+}+\Delta r}^{r_{\rm max}} \frac{r^{2}+a^{2}M^2}{r^{2}-2Mr+a^{2}M^2} dr  \nonumber \\
=2r_{\rm max}-2r_{+}-2\Delta r + 2\frac{r_{+}^{2}+a^{2}M^2}{r_{+}-r_{-}} \ln(\frac{r_{\rm max}-r_{+}}{\Delta r})  \nonumber \\
- 2\frac{r_{-}^{2}+a^{2}M^2}{r_{+}-r_{-}} \ln(\frac{r_{\rm max}-r_{-}}{r_{+}-r_{-}+\Delta r}), \ \ \ \ \ \ \ \ \ \ \ \ 
\end{eqnarray}
where $r_{\pm}=M(1\pm \sqrt{1-a^{2}})$,  and $\Delta r$ is the coordinate distance of the membrane and the (would-be) horizon.

The peak of the angular momentum barrier, $r_{\rm max}$, is given by the roots of a sixth-order polynomial \cite{Yang:2013uba}:
\begin{eqnarray}
(1-\mu^{2})[(2-\mu^{2})\hat{r}_{\rm max}^{2} + 2(2+\mu^{2})\hat{r}_{\rm max} + (2-\mu^{2})]a^{4} \nonumber \\
 + 4 \hat{r}_{\rm max}^{2} [(1-\mu^{2})\hat{r}_{\rm max}^{2} - 2 \hat{r}_{\rm max} -3 (1-\mu^{2})]a^{2} \ \ \ \ \ \ \ \ \ \nonumber \\
+2 \hat{r}_{\rm max}^{4}(\hat{r}_{\rm max}-3)^{2}=0, \ \ \ \ \ \ \ \ \ \ \ \ \ \ \ \ \ \ \ \ \ \ \ \ \ \ \ \ \ \ \ \ \ 
\end{eqnarray}
where $\mu=m/(l+\frac{1}{2})$ and $\hat{r}_{\rm max} =r_{\rm max}/M$. For the dominant QNM, $r_{\rm max} < 3 M$ and $(l,m)=(2,2)$ resulting in  $\mu=0.8$.

We further posit that the location of the membrane should be near a Planck proper length from the horizon. This assumption is required to explain the observed density of cosmological dark energy within the gravitational aether proposal \cite{PrescodWeinstein:2009mp}, but is also expected from generic quantum gravity scalings, such as the brick wall model \cite{THOOFT1985727}, or trans-Planckian effects \cite{Khriplovich,PhysRevD.28.2929}. This implies:
\begin{eqnarray}
\int_{r_{+}}^{r_{+}+\Delta r} \sqrt{g_{rr}} dr |_{\theta=0} \sim l_p \simeq 1.62 \times 10^{-33}~ {\rm cm},
\end{eqnarray}
which fixes the location of the membrane:
\begin{eqnarray}
\Delta r|_{\theta=0}=\frac{\sqrt{1-a^{2}} l_{p}^{2}}{4M(1+\sqrt{1-a^{2}})}.
\end{eqnarray}

With this set-up, we note that $\Delta t_{\rm echo}\simeq 8M\log(M/l_p)\left[1+{\cal O}(a^2)\right]$ is comparable to the scrambling time: the time over which the black hole state is expected to thermalize \cite{Hayden:2007cs,Sekino:2008he,Harlow:2014yka,Harlow:2013tf}.

Using the measurements of the final black hole (redshifted) mass and spin by the LIGO collaboration, we can constrain $\Delta t_{\rm echo}$ for each merger event. Assuming gaussian errors, we find (see Appendix D for details of calculations):
\begin{eqnarray}
\boxed{\Delta t_{{\rm echo}, I }({\rm sec})
=\left\{
 \begin{matrix}
  0.2925 \pm 0.00916 & I= {\rm GW150914} \\
0.1013 \pm 0.01152 & I={\rm GW151226} \\
 0.1778 \pm 0.02789 & I={\rm LVT151012}
 \end{matrix}
 \right.} \ \ \ \ \ \label{t_echo_meas}
 \end{eqnarray}



\section{ Data and the Echo template} \label{sec_template}

In this analysis, we use four datasets for each event. The first two are the theoretical best-fit waveform for Hanford and Livingston detectors (in real time series) for the BH merger event, provided by the  LIGO and Virgo collaborations \cite{GW150914,GW151226,LVT151012,Vallisneri:2014vxa}. The other two are the observed strain datastream from the two detectors. We call these $M_{H,I}(t)$, $M_{L,I}(t)$, $h_{H,I}(t)$ and $h_{L,I}(t)$, respectively.  We used the strain data at 4096 Hz and for 32 sec duration. The waveform model consists of three phases: inspiral, merger, and ringdown. 

Following the numerical results of \cite{Cardoso:2016rao,Cardoso:2016oxy}, we construct a phenomenological gravitational wave template for the echoes using five free parameters:
\begin{enumerate}
\item $\Delta t_{\rm echo}$ is the time-interval in between successive echoes, which we vary within the $1\sigma$ range, fixed by the uncertainties in (redshifted) mass and angular momentum of the final black hole (Eq. \ref{t_echo_meas}). 

\item $t_{\rm echo}$ is the time of arrival of the first echo, which can be affected by non-linear dynamics near merger (Eq. \ref{nonlinear}).

\item $t_0$ quantifies which part of the GR merger template is truncated to produce the subsequent echo templates\footnote{Note that the wavelength of gravitational waves in the inspiral phase is much longer than the size of the black holes, which leads to an echo signal suppressed at 4PN order \cite{Poisson:1994yf}.}. To do this, we introduce a smooth cut-off function with a free parameter $t_{0}$,
\begin{equation}
\Theta_I(t, t_{0})\equiv\frac{1}{2}\left\{1+ \tanh\left[\frac{1}{2} \omega_I(t)(t-t_{\rm merger}-t_{0})\right]  \right\},
\end{equation}
where $\omega_I(t)$ is frequency of model as a function of time \cite{TheLIGOScientific:2016src} and $t_{\rm{merger}}$ is the time at which the GR template peaks. As the intermediate region (merger) is before $t_{\rm{merger}}$, we assume $t_{0}$ is negative, and vary it within the range $t_{0} \in (-0.1,0) \overline{\Delta t}_{\rm echo}$. Using this definition, we can define the truncated template:
\begin{eqnarray}
{\cal M}_{T,I}^{H/L} (t, t_{0}) \equiv\Theta_I(t, t_{0}) {\cal M}_{I}^{H/L} (t).
\end{eqnarray}

\item $\gamma$ is damping factor of successive echoes, which should be between $0$ and $1$. In our analysis,  we vary this free parameter within the range $(0.1,0.9)$ and look for the best fit.

\item $A$ is the over-all amplitude of the echo template (with respect to the main event) which we fit for, assuming a flat prior. 

\end{enumerate}

\begin{figure}[!tbp]
  \centering
  \begin{minipage}[b]{0.5\textwidth}
    \includegraphics[width=1\textwidth]{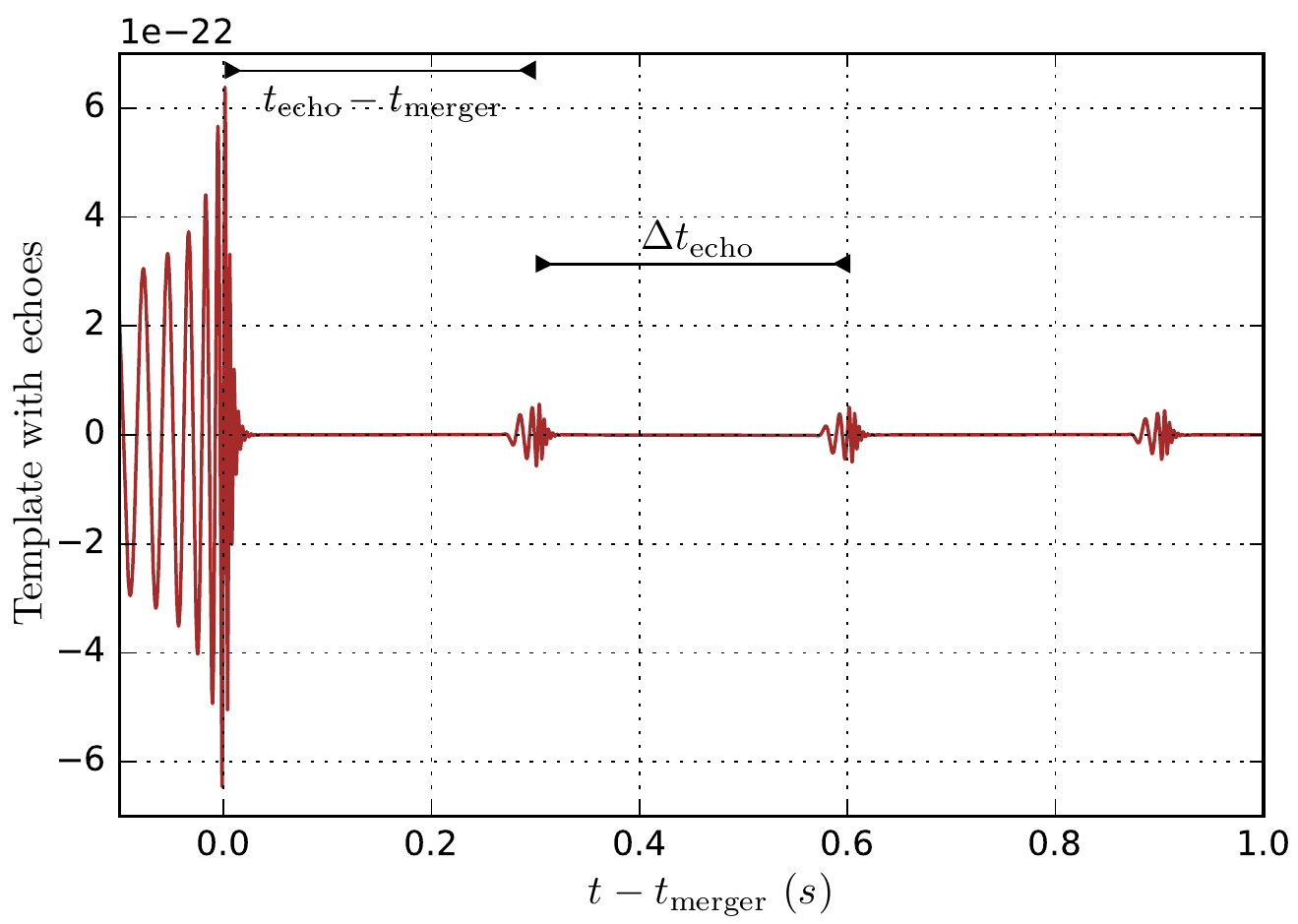}
  \end{minipage}
\caption{LIGO original template for GW150914, along with our best fit template for the echoes.\label{template_echoes}}
\end{figure}

\begin{figure}[!tbp]
  \centering
  \begin{minipage}[b]{0.5\textwidth}
    \includegraphics[width=1.1\textwidth]{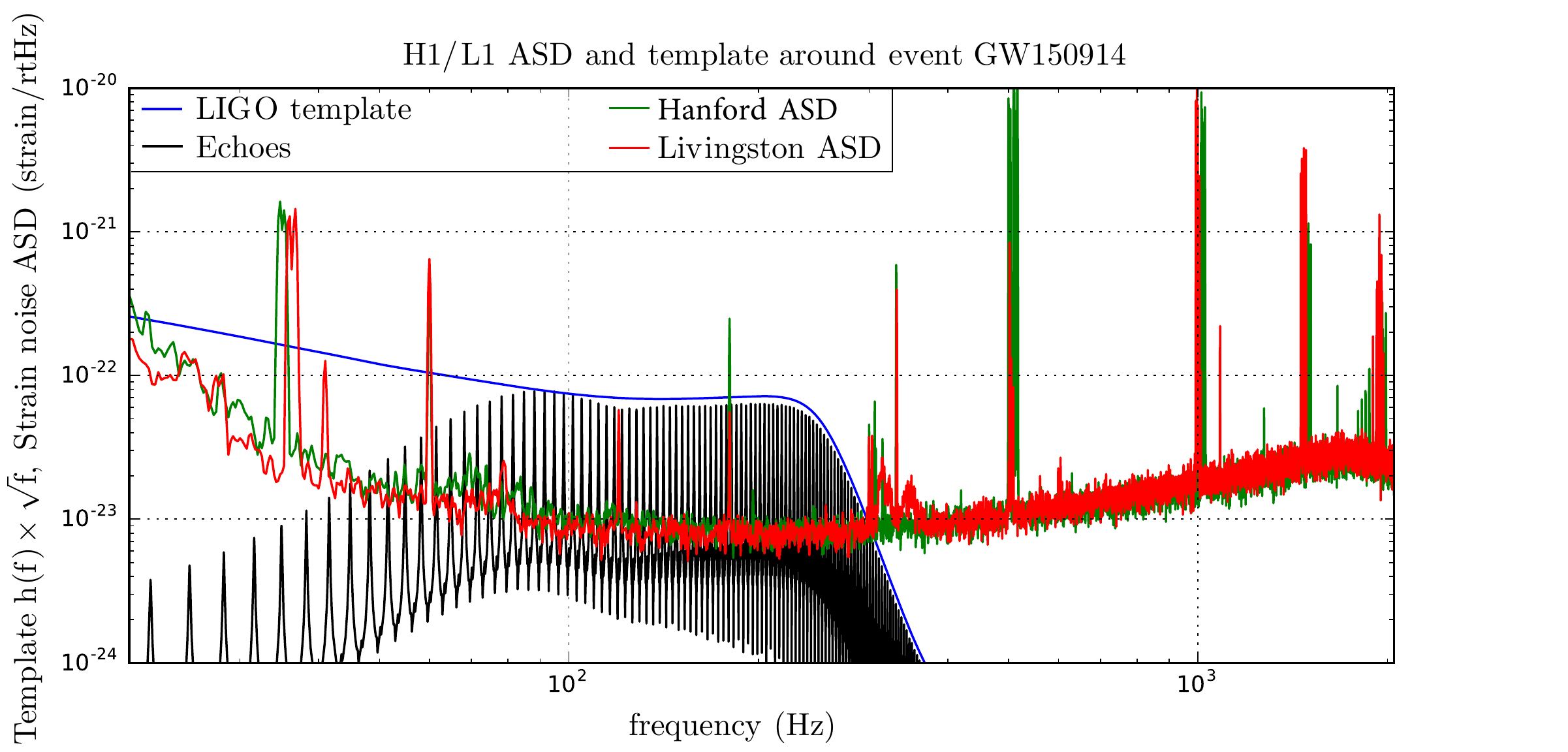}
  \end{minipage}
\caption{Amplitude Spectral Densities (ASD's) of our best fit echo template (Eq. \ref{template}) and the main event, for GW150914. Since we have a quasi-periodic model, there are resonances in the spectrum. The ASDs are the square root of the power spectral densities, which are averages of the square of the fast Fourier transforms of the data. The noise spectra from Hanford and Livingston detectors are also shown.}\label{template_matchfreq}
\end{figure}

The truncated model with echoes and all the free parameters is then given by:
\begin{eqnarray}
\!\!\!\! &&M_{TE,I}^{H/L}(t) \equiv \nonumber\\
\!\! && A\displaystyle\sum_{n=0}^{\infty}(-1)^{n+1}\gamma^{n} {\cal M}_{T,I}^{H/L}(t+t_{\rm merger}-t_{\rm echo}-n\Delta t_{\rm echo},t_{0})\ \ \ \ \ \label{template}
\end{eqnarray}
where the term $(-1)^{n+1}$ is due to the phase inversion of the truncated model in each reflection. Fig. (\ref{template_echoes}) shows our best fit for this template for GW150914 within the parameter space described above, along with the main merger event. In the frequency domain we expect to see resonances of these echoes (Fig. \ref{template_matchfreq}). 

\begin{figure}[!tbp]
    \includegraphics[width=0.5\textwidth]{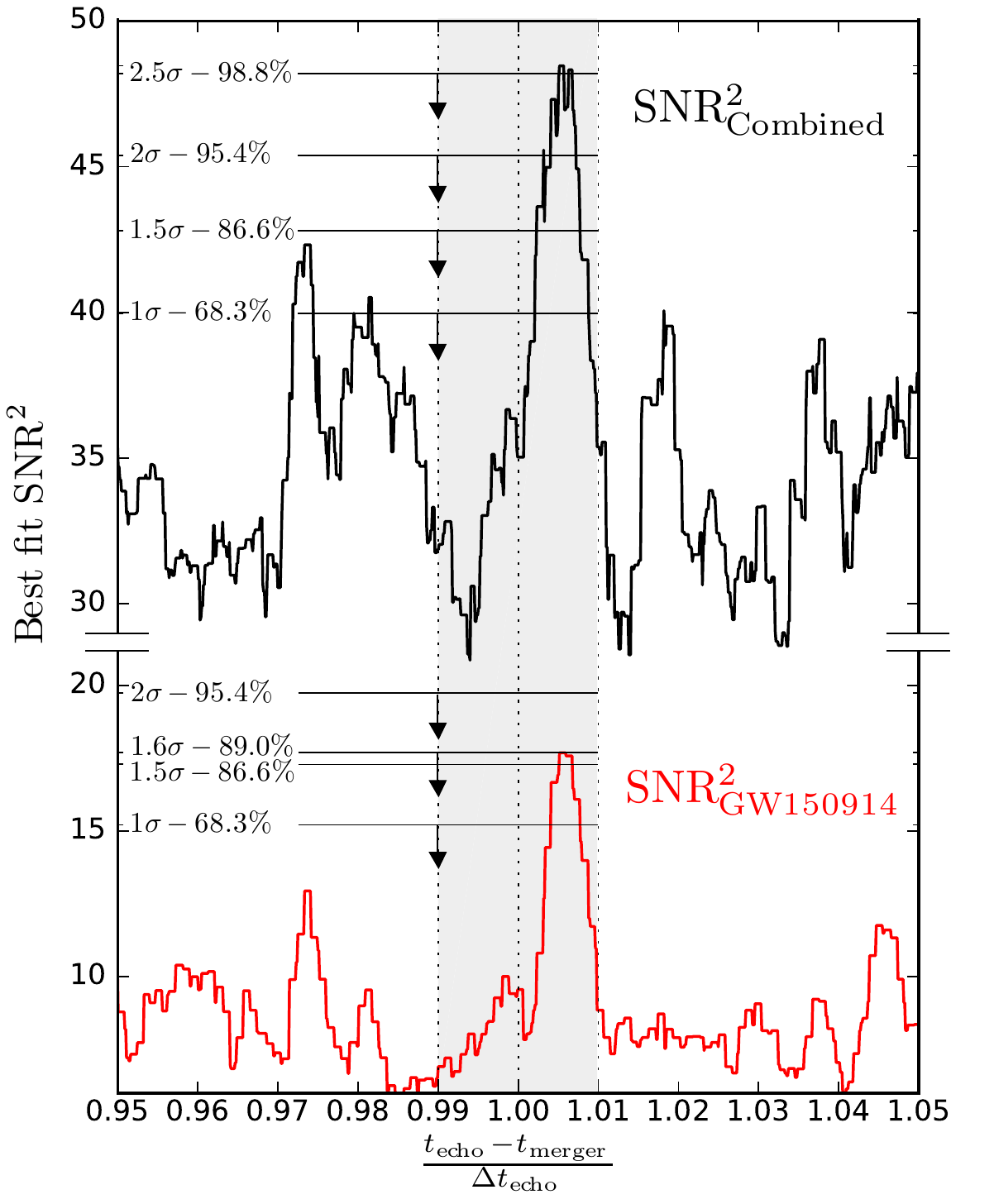}
 \caption{Best fit (or maximum) SNR$^2$ near the expected time of merger echoes (Eq's. \ref{nonlinear} and \ref{t_echo_meas}), for the combined (top) and GW150914 (bottom) events. The significance of the peaks is quantified by the p-value of their SNR$_{\rm max}$ within the gray rectangle (see Appendix E for detail of calculation).}\label{SNR}
\end{figure}

\begin{figure}[!tbp]
    \includegraphics[width=0.5\textwidth]{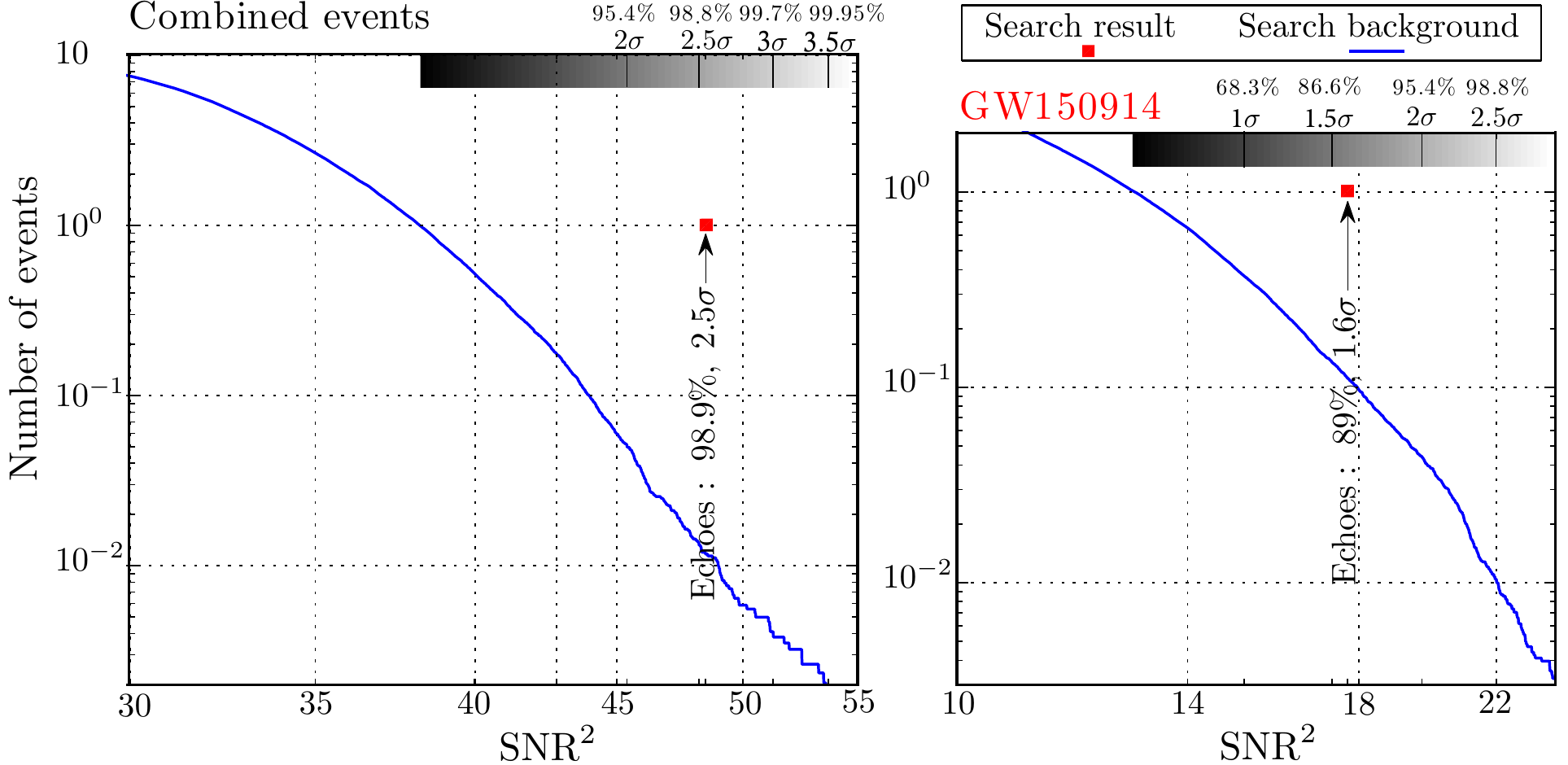}
    \caption{Average number of noise peaks higher than a particular SNR-value within a  time-interval $2\% \times   \overline{\Delta t}_{\rm echo}$ for combined (left) and GW150914 (right) events.  The red dots show the observed SNR peak at $t_{\rm echo} = 1.0054 \Delta t_{\rm echo}$ (Fig. \ref{SNR}). The horizontal bar shows the correspondence between SNR values and their significance.}    
 \label{Histogramloglog}
\end{figure}

\begin{table}
\begin{center}
\begin{tabular}{ |c|c|c|c|c| }
\hline
 & Range& GW150914& Combined  \\
\hline
$(t_{\rm echo} -t_{\rm merger})/ \Delta t_{\rm echo}$& (0.99,1.01)& 1.0054& 1.0054  \\
$\gamma$& (0.1,0.9)& 0.89& 0.9 \\
$t_{0}/\overline{\Delta t}_{\rm echo}$& (-0.1,0)& -0.084& -0.1 \\
Amplitude\footnote{The combined amplitude is given by: $A_{\rm{average}}=\frac{\displaystyle\sum_{I}\frac{SNR_{I}^{2}}{|A_{I}|}}{\displaystyle\sum_{I}\frac{SNR_{I}^{2}}{A_{I}^{2}} }$}&   &0.0992 &0.124 \\
SNR$_{\rm max}$  &   & 4.21 & 6.96 \\
\hline
p-value &  &    $0.11$ &$0.011$ \\
significance &  & 1.6$\sigma$ & 2.5$\sigma$\\
\hline
\end{tabular}
\caption{Best fit values for echo parameters of the highest SNR peak near the predicted $\Delta t_{\rm echo}$, and their significance.   }\label{table_1}
\end{center}
\begin{center}
\begin{tabular}{ |c|c|c|c| }
\hline
\ \  & GW150914 & GW151226 & LVT151012 \\
\hline
$\Delta t_{\rm echo, pred}$(sec) & 0.2925  &  0.1013  & 0.1778 \\
 & $\pm$ 0.00916 & $\pm$ 0.01152 & $\pm$ 0.02789 \\
\hline
$\Delta t_{\rm echo, best}$(sec) & 0.30068 & 0.09758 & 0.19043 \\ 
$|A_{\rm best, I}|$ & 0.091 & 0.33 & 0.34 \\ 
SNR$_{\rm best, I}$ & 4.13 & 3.83 & 4.52\\
\hline
\end{tabular}
\caption{Theoretical expectations for $\Delta t_{\rm echo}$'s of each merger event (Eq. \ref{t_echo_meas}), compared to their best combined fit within the 1$\sigma$ credible region, and the contribution of each event to the joint SNR for the echoes (Eq. \ref{snr_total}).
}\label{table_2}
\end{center}
\end{table}

\section{Results}

 Our strategy is to search for the best fit for the echo template (\ref{template}), by maximizing its signal-to-noise ratio, SNR, within the conjectured parameter space described above at fixed $x=(t_{echo}-t_{merger})/\Delta t_{echo}$. We then identify the highest peak within the range $t_{\rm echo} - t_{\rm merger} = \left(1 \pm 0.01\right)\Delta t_{\rm echo}$ (Eq. \ref{nonlinear}). This range in $t_{\rm echo}$ is expected, e.g., due to a random phase in the complex echo template (see \cite{Abedi:2017isz} and Appendix C). We quantify the significance of this peak by how often a higher SNR peak is achieved within an interval of duration  $2 \% \times \overline{\Delta t}_{{\rm echo}, I}$, in the background (away from the main event) in the data stream, where $\overline{\Delta t}_{{\rm echo}, I}$ is the mean of $\Delta t_{{\rm echo}, I }$ for independent events in Eq. \ref{t_echo_meas}. 
It is worth noting that due to different angles and locations of each detector, a complex model is analysed. Therefore in calculation of SNR's we subtracted the phase of the main event from complex template and obtained two real templates corresponding to each detector (Hanford/Livingston). Then we set the original gravitational wave peak at $t=0$ by removing the offset from SNRs (see \cite{GW150914,GW151226,LVT151012,Vallisneri:2014vxa}).
 We combine the SNR's of different detectors for each event by adding the $\chi^2$ for two datasets, using the same echo model\footnote{The proper way of combining datasets is through inverse weighting by $\rm{Noise}^{2}$. We use the analysis packages provided by the LIGO Open Science Center (https://losc.ligo.org) to compute SNR for a given template and combine it for two detectors.}.

We do the analysis once for GW150914 (LIGO's most significant detection), and repeat it for the 3 recorded events combined, by maximizing:
\begin{eqnarray}
 {\rm SNR}^2_{total} \equiv \sum_I {\rm SNR}^2_I. \label{snr_total}
 \end{eqnarray}
 In doing so, we assume the same $\gamma$ and $t_0/\overline{\Delta t}_{\rm echo}$ for all three events, while keeping $\Delta t_{\rm echo}$ and $A$'s independent. The results are shown in Fig's (\ref{SNR}-\ref{Histogramloglog}) and Tables \ref{table_1}-\ref{table_2}.

Fig. (\ref{SNR}) shows that there is indeed a significant peak with SNR$_{\rm max}= 4.21 (6.96)$ for echoes following the GW150914 (combined) merger event(s), within 0.54\% of the predicted echo time delay. To find the significance of finding this peak so close to the predicted value, we divide up the data steam (within the range 9-38 $\times  \overline{\Delta t}_{\rm{echo},I}$ after the merger) into intervals of $2 \% \times \overline{\Delta t}_{{\rm echo}, I}$, and compute the average number of points in the interval that exceed SNR (Fig.   \ref{Histogramloglog}). 
 This yields an estimate of the false detection probability and the significance of SNR peaks observed near the predicted echo times, at 0.11 (0.011) and 1.6$\sigma$ (2.5$\sigma$) for  the GW150914 (combined) merger event(s) respectively. More discussion of our statistical methodology, and possible alternatives, can be found in Appendix and \cite{Abedi:2017isz,Ashton:2016xff}.


\section{Conclusions and Discussion} 

In this {\it paper}, we have searched advanced LIGO's  public data release of the first observed gravitational wave signals from black hole merger events for signatures of Planck-scale structure near their event horizons. By building a phenomenological template for successive echoes from such exotic structures expected in e.g., {\it firewall} or {\it fuzzball} paradigms, after marginalizing over its parameters, we report a first tentative evidence for these echoes at false detection probability of 0.011 or $2.5\sigma$ significance in LIGO data proceeding its reported merger events. Future data releases from LIGO collaboration at higher sensitivity will be able to definitively confirm or rule out this finding.

One may wonder how including GW151226 and LVT151012 may improve the significance of the echo signal, even though their ringdown phase was not detectable in LIGO data. We should note that while GW150914 and the two other events have similar numerical contributions to the significance, the nature of their contributions are quite different: GW150914 appears to pin down the echo template parameters, while the others help improve evidence for this template. Furthermore, the repeating nature of the echoes gives them a low frequency structure (Fig. \ref{template_matchfreq}) which may be detectable, even if the ringdown itself falls below the detector noise at high frequencies for GW151226 and LVT151012 (Fig. 1 in  \cite{TheLIGOScientific:2016pea,TheLIGOScientific:2016agk}).

We should note that the {\it ad hoc} nature of our echo template construction is not entirely satisfactory and could lead to some ambiguity in interpreting the statistical significance of our finding. In particular, the fact that the combined SNR is maximized on the edge of our parameter range (see Table \ref{table_1}) points to a need for a better physical prior on parameters, or simply a more physical echo template. For instance, increasing the range of prior for $t_{0}$ adds a significant portion of the inspiral into the echo template, which may suggest a need for a less {\it ad hoc} truncation function. 
This does not change the statistical significance of our SNR peaks, but suggests better fits may lie beyond this range (see Appendix for more discussions). In addition, reliable extension of the analysis beyond this range (in particular $\gamma > 0.9$) requires analyzing a much larger portion of LIGO data, where one may also worry about the time variability and non-gaussianity of the LIGO instrumental noise (see Fig's 14-15 in \cite{Martynov:2016fzi} and Appendix).

From a physical standpoint, a slowly damping echo, $\gamma \approx 1$, may not be unexpected and could be intimately related to the well-established instability of horizonless ergoregions \cite{1978CMaPh..63..243F,Cardoso:2007az}.  Future numerical simulations of merging black holes with a membrane can sharpen the echo template, possibly increasing the detection significance. We thus predict that a synergy of improvements in observational sensitivity and theoretical modelling can provide conclusive evidence for quantum gravitational alternatives to black hole horizons.

\begin{center} {\bf Note Added} \end{center}

Since we submitted this paper, several authors have proposed alternative templates and/or search methodologies for ``echoes from the abyss''.  For example, \cite{Mark:2017dnq} used a Green's function method to find scalar echoes generated by a point mass on a marginally unstable circular orbit around a Schwarzschild black hole surrounded by a (partially) reflective wall.  In contrast, \cite{Nakano:2017fvh} used the reflectivity of the angular momentum barrier for a Kerr black hole to construct a phenomenological echo template. While these treatments included more realistic elements of wave propagation (compared to our approach), they still miss two crucial pieces of physics, namely a frequency-dependent reflectivity for the fuzzball/firewall (see \ref{Introduction} above), and realistic initial conditions that should ideally come from non-linear simulations of binary mergers.  Therefore, it is not clear whether these templates are significantly more realistic than a phenomenological model such as the one adopted in \ref{sec_template} above. 

Similar to our approach, Maselli {\it et al.} \cite{Maselli:2017tfq} provide a phenomenological echo template based on superposition of sine-gaussians with free parameters, and forecast how well they can be measured using Advanced LIGO observations. It will be interesting to see the application of this method to real data, and whether it can recover similar tentative evidence for echoes.

\begin{acknowledgments}

We  thank Vitor Cardoso, Will East, Sabine Hossenfelder, Matt Johnson,  Luis Lehner, Rafael Sorkin, Nico Yunes, Qingwen Wang, and Aaron Zimmerman for helpful comments and discussions. 
We especially thank the authors of \cite{Ashton:2016xff} for their constructive criticism, as well as the anonymous referees for the valuable comments and suggestions to improve the clarity and the quality of this paper.
 We also thank all the participants in our weekly group meetings for their patience during our discussions, as well as Jonah Miller and Erik Schnetter for computational help in this project. JA would like to thank Perimeter Institute for Theoretical Physics (PI) for the great hospitality during the course of this work on his collaboration leave, as well as his research adviser Hessamaddin Arfaei for all his supports. He also thanks Mansour Karami, Seyed Faroogh Moosavian and Yasaman Yazdi for their kind hospitality during his visit. 
This research has made use of data, software and web tools obtained from the LIGO Open Science Center (https://losc.ligo.org), a service of LIGO Laboratory and the LIGO Scientific Collaboration. LIGO is funded by the U.S. National Science Foundation. 
This work has been partially supported by Ministry of Science, Research and Technology of Iran, Institute for Research in Fundamental Sciences (IPM), University of Waterloo, and Perimeter Institute for Theoretical Physics (PI). Research at PI is supported by the 
Government of Canada through the Department of Innovation, Science and Economic Development Canada and by the Province of Ontario through the Ministry of Research, Innovation and Science.

\end{acknowledgments}

\appendix
\section{Estimation of the total echo energy emission}

With a simple assumption we may be able to estimate the total energy emitted in echoes. Due to physical reasons the inspiral part is removed and each echo has energy equivalent to $\gamma^{2n} E_{\rm{Ringdown\ Merger}}=\gamma^{2n} \xi E_{Total}$. Where $0<\xi <1$ is the fraction of the energy in ringdown and merger part. With this given assumption we obtain,
\begin{eqnarray}
\frac{E^{I}_{\rm{echoes}}}{E^{I}_{\rm{Ringdown\ Merger}}}=\frac{A_{I}^{2}}{1-\gamma^{2}}.
\end{eqnarray}
Since the inspiral part for different events are not identical, the portion of the energy in inspiral part varies. Hence, we shall obtain different values of $\xi_{I}$ for different events. Here we have following best fit values for different events,
\begin{eqnarray}
\xi_{I} \simeq \left\{
 \begin{matrix}
  0.24 & I=GW150914 \\
  0.16 & I=GW151226 \\
  0.17 & I=LVT151012
 \end{matrix}
 \right.\nonumber \\ \label{angle}
\end{eqnarray}

\begin{figure}[!tbp]
    \includegraphics[width=0.5\textwidth]{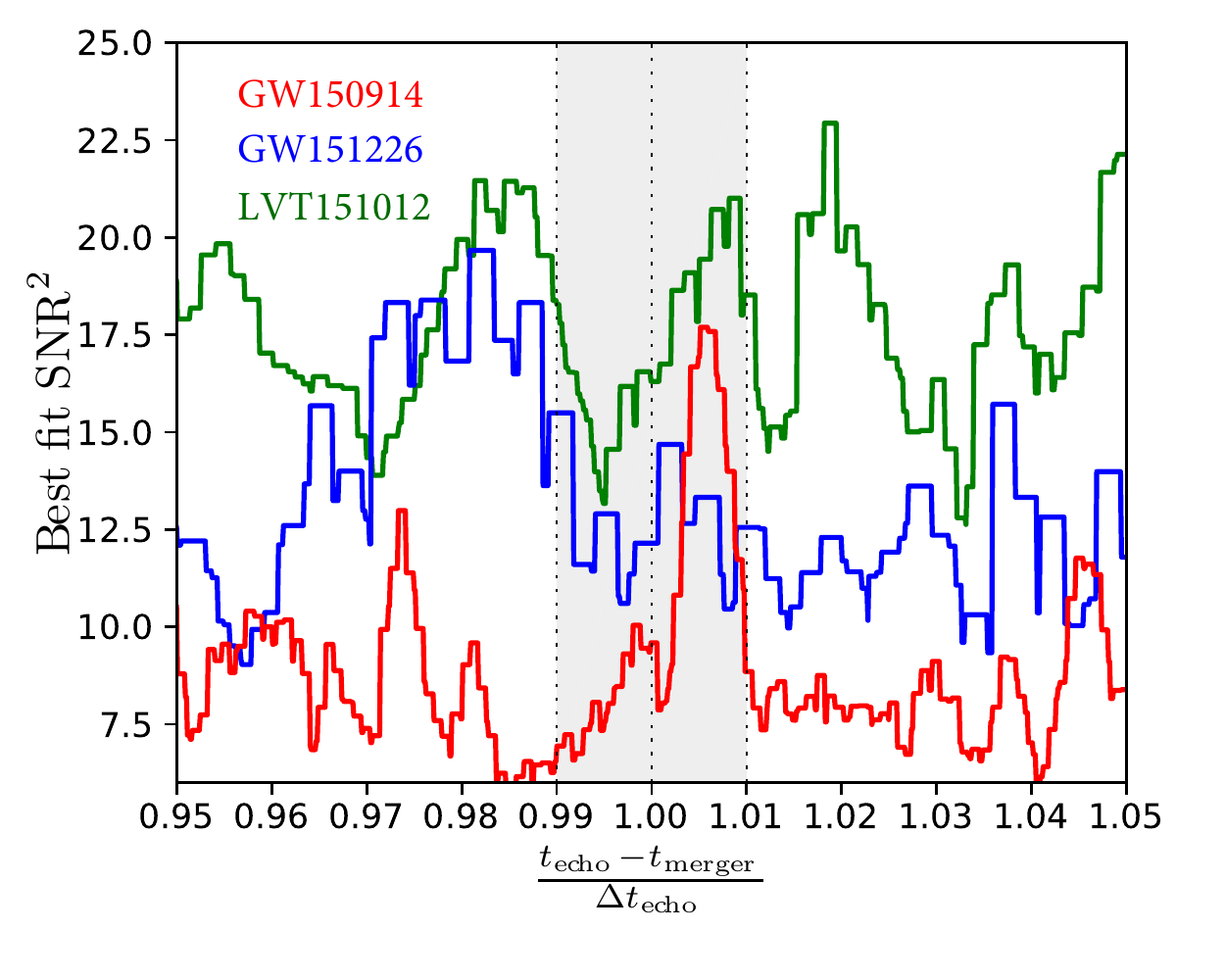}
 \caption{Best fit (or maximum) SNR$^2$ near the expected time of merger echoes (similar to the Fig. \ref{SNR} in the main text), for all the three events.}\label{Three_events}
\end{figure}

Therefore, the total echoes energy emission is given as follows,

\begin{eqnarray}
E^{I}_{\rm{echoes}}/(M_{\odot}c^{2})=\left\{
 \begin{matrix}
  0.029 & I=GW150914, \\
  0.047 & I=GW151226, \\
  0.16 & I=LVT151012,
 \end{matrix}
 \right.\nonumber \\ \label{angle}
\end{eqnarray}
which are the energies emitted in the source frame.
One important point to consider is that since GW151226 and LVT151012 contribute to the significance differently, these results for energy emission may not be valid for them. As it is shown in Fig. \ref{Three_events} the background for GW151226 and LVT151012 is relatively higher than GW150914. Therefore, the amplitude used in this calculation is affected by background, systematically biasing the energy to higher values, especially for lower significance events.

\section{Detection properties of echo template}
Is the sensitivity of LIGO detectors enough to probe echoes from the ringdown in the data? In Fig. \ref{Match_frequency}, we see that the {\it best-fit} echo templates are mostly concentrated around the minimum of the amplitude spectral distribution (ASD) of data, where the detector has highest sensitivity. In addition, due to the cumulative nature of echoes, they produce resonance peaks comparable to the amplitude of the main event even for echoes with overall amplitude 10 times lower than the original event (see Fig. \ref{Match_frequency}). 

Given their best fit values in Table \ref{table_1}, one may wonder whether the prior ranges on the parameters $\gamma$ and $t_{0}$ are too narrow. 
However, changing the prior for these parameters based on their best-fit values leads to {\it a posteriori} statistics and can bias the p-values. Furthermore, allowing the noise statistics to drive the priors for the model parameters, we may end up with an unphysical range that adversely affect echo searches in the future data releases.

For the damping factor $\gamma=0.9$ at the boundary of our prior range, one may worry that it might pose a problem for our analysis (an issue that we discussed at length in the main text) . This indeed would be the case if the goal was to measure these parameters. However, that has not been our goal, as the parameters only quantify a toy model for the echoes. The  goal was rather to find whether the best-fit toy model, within the parameter range, is consistent with random noise. As we discussed in the main text, we find that has a probability of $<$ 1\%. 

\begin{figure*}
     \includegraphics[width=0.45\textwidth]{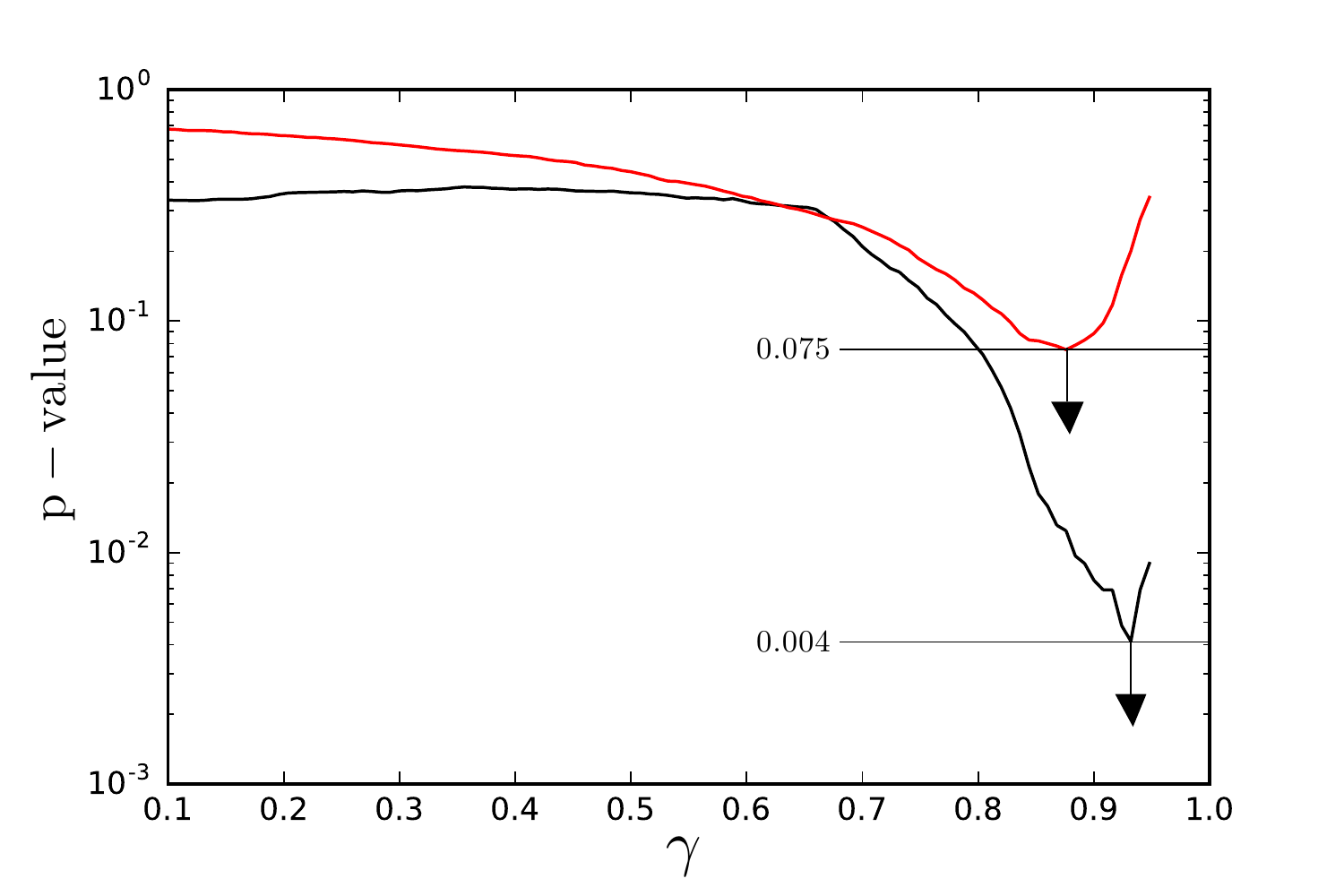}
     \includegraphics[width=0.5\textwidth]{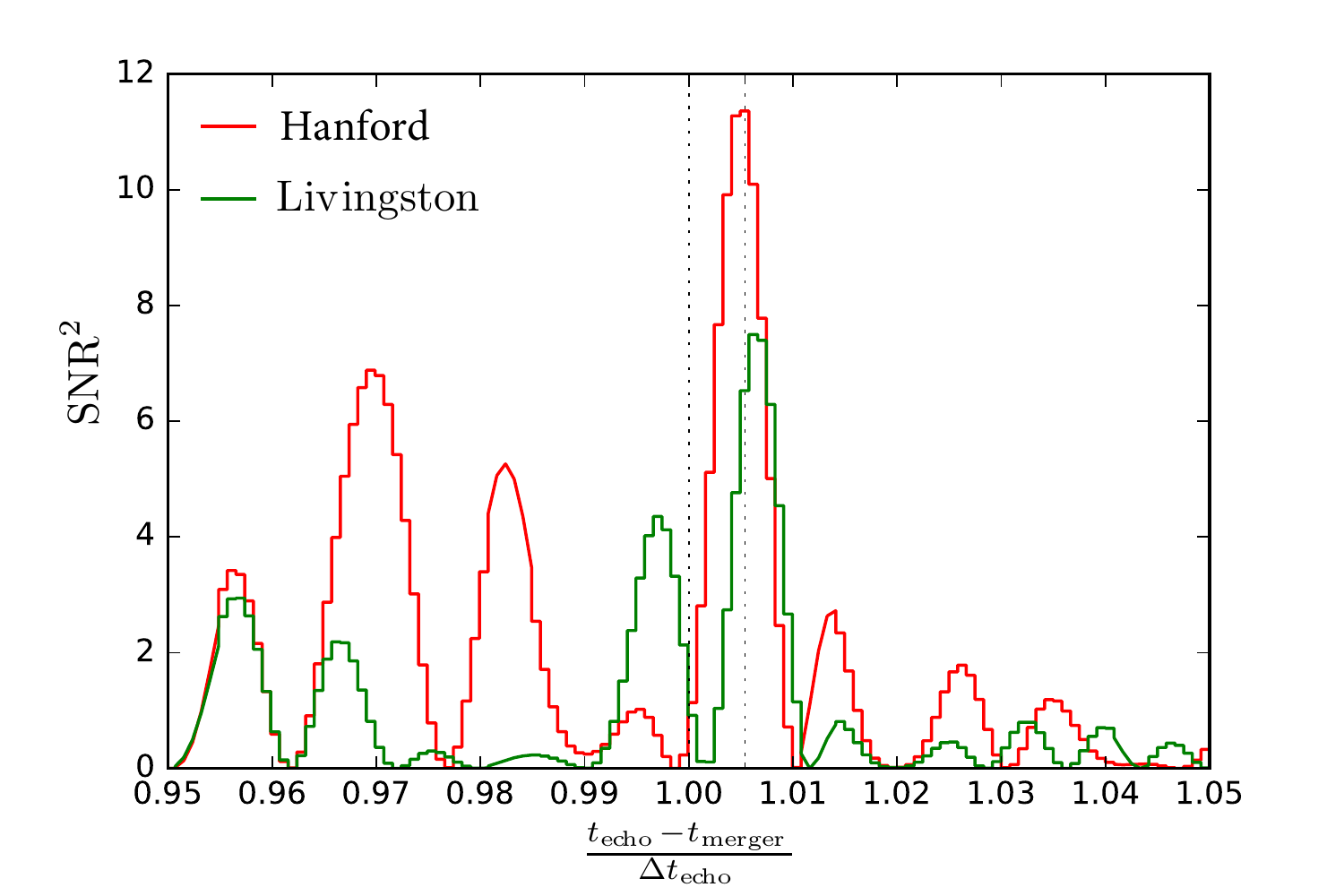}
    \caption{(a) {\it (Left)}  p-value for our maximum SNR, as a function of $\gamma$, computed within a time-interval $ (-1\%,1\%) \times   \overline{\Delta t}_{\rm echo}$ for GW150914 (top) and combined events (bottom).
\\
  (b) {\it (Right)} SNR$^{2}$ near the expected time of best-fit merger echoes (Eq. \ref{t_echo_meas}
) for GW150914 in Hanford (red) and Livingston (green) detectors. Not only do we see that the two detectors see coincident SNR peaks,  but also their ratio $2.42/3.49 = 0.69$ is comparable to the SNR ratio for the main events $13.3/18.6 = 0.72$ seen in the two detectors. Note that, unlike Fig. \ref{SNR} in the main text,
 here we have fixed the echo parameters to their best fit values for combined detectors \cite{Abedi:2017isz}.
}
 \label{Hanford_Livingston}
\end{figure*}

Fig. (\ref{Hanford_Livingston}a) shows p-value as a function of the echo damping factor $\gamma$. We can see that for less damped echoes, p-value drops significantly, which provides substantial evidence for the existence of echoes of gravitational wave in the LIGO data. Here, p-value as a function of $\gamma$ yields a p-value of 0.004 or a maximum significance of 2.9$\sigma$ at $\gamma=0.93$ for combined events. However, we should caution that, as we discuss in the main text, given the duration of data used in the analysis, the interpretation becomes less reliable for $\gamma >0.9$.  Furthermore, Fig. (\ref{Hanford_Livingston}a) shows that the p-value has a jump at $\gamma>0.9$, signifying a jump in the best fit. This can be understood by noticing that as $\gamma$ increases, the length of the template increases, eventually diverging in the limit $\gamma \rightarrow 1$. This is a singular limit, as arbitrarily high SNR's can be found by simply fitting the noise. Therefore, our method of maximizing SNR fails in this limit, and $\gamma \sim 1$ should be avoided in the prior.   

As we argue in the main text, rather than pushing the parameters of a toy model to their extremes, in our opinion, it will be much more fruitful to find more physical echo templates, an effort that is already underway by several research groups.

Finally, one may wonder why combining events with poor individual evidence for echoes can strengthen the significance of the claimed echoes for GW150914. Here, the rationale is that in our model, each echo is quantified by 5 parameters that are poorly constrained. Given the marginal nature of the signal, there will be large degeneracies amongst these parameters from individual events. However, since echoes for different events have different frequency coverage (Fig. \ref{Match_frequency}), the data constrains different combinations of the parameters. Therefore, they can be combined to (at least partially) break these degeneracies and reduce the error on amplitudes leading to a (marginal) detection.

\begin{figure*}
  \centering
    \includegraphics[width=\textwidth]{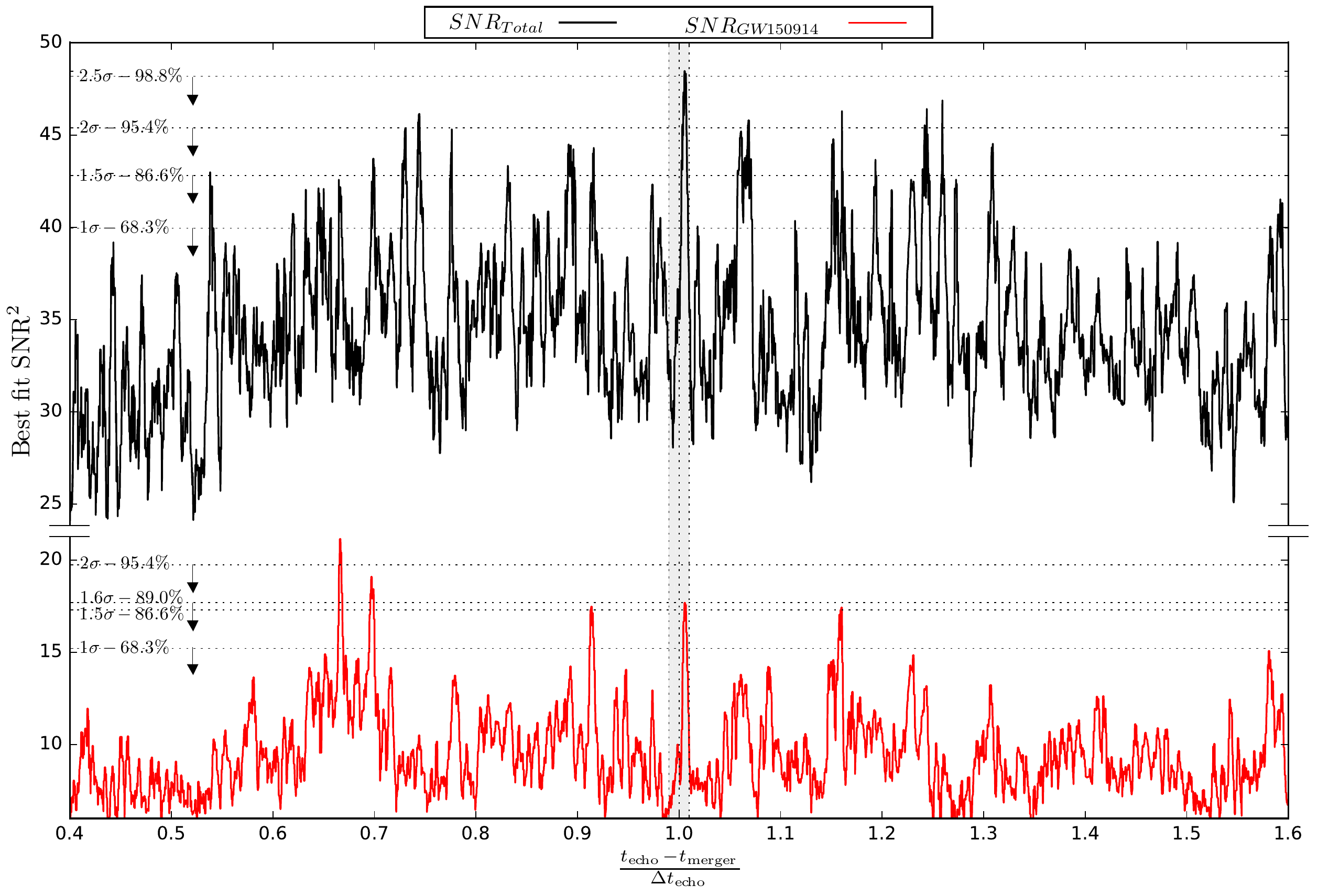}
    \caption{Same as Fig. \ref{SNR} 
    in the main text, but over an extended range of $x=\frac{t_{\rm echo} - t_{\rm merger}}{\Delta t_{\rm echo}}$. The SNR peaks at the predicted value of $x=1$ have false detection probability of 0.11 (0.011) and significance of 1.6$\sigma$ (2.5$\sigma$), for GW150914 (combined events)  (See also \cite{Abedi:2017isz}). } \label{SNR_fig}
\end{figure*}


\section{The Holiday Edition!}

While we report tentative evidence for the presence of echoes from Planck-scale modifications of general relativity, our statistical methodology was challenged by Ashton, et al. \cite{Ashton:2016xff}. In this section, we summarize our response \cite{Abedi:2017isz} which addresses these criticisms.

\begin {enumerate}

\item Ashton et al. point out that we find a slightly higher SNR$_{\rm best}$ for echoes in LVT151012, compared to GW150914, even though the SNR for the main event is lower by a factor of 2.4.  Is this surprising?
In fact, this is expected as constraints on final mass and spin of LVT151012 are significantly worse than GW150914. As a result, the relative error on $\Delta t_{\rm echo}$ is 5 times higher for LVT151012, compared to GW150914. This leads to larger values of SNR$_{\rm best}$ across the board, as we are searching a larger region of parameter space. This, however, does not necessarily lead to increased significance, as the same would be true for all values of $x=\frac{t_{\rm{echo}}-t_{\rm{merger}}}{\Delta t_{\rm{echo}}}$.

If there was no real echo signal in LVT151012 and GW151226, adding them to GW150914 would only dilute the significance of the peak near $x=1$. The fact that the opposite happens suggests that, in spite of larger variations in SNR  due to higher uncertainty in $\Delta t_{\rm echo}$, there is still significant enhancement in SNR near $x=1$.

We should also caution about comparing the significance of the echoes with that of the merger events, as they have very different frequency structures (see Fig. \ref{Match_frequency}) leading to different SNR ratios, especially given the non-trivial frequency dependence of the LIGO detector noise.

Finally, we should warn about over-interpreting our quoted significances. Even though we gain comparable evidence for echoes by including LVT151012 and GW151226, i.e. $ 1.6^2+1.6^2 \simeq 2.5^2$, it doesn't mean that they have the same significance: A $1.6\sigma$ peak could be a 1-$\sigma$ fluctuation of a 0.6-$\sigma$ or a 2.6-$\sigma$ underlying signal.

For completeness, the individual amplitudes of the best joint fit are listed in Table \ref{table_2} in the main text. We note that, even though best fit SNR's are comparable for the three events, the errors on the amplitude: $\Delta A$ = $A_{\rm best}$/SNR$_{\rm best}$ is much smaller for GW150914, given that $A_{\rm best}$ is the smallest. Therefore, as expected, GW150914 which is the most significant of the 3 LIGO events, would also dominate the combined constraint on the echo amplitude.

\begin{figure}
  \centering
    \includegraphics[width=0.5\textwidth]{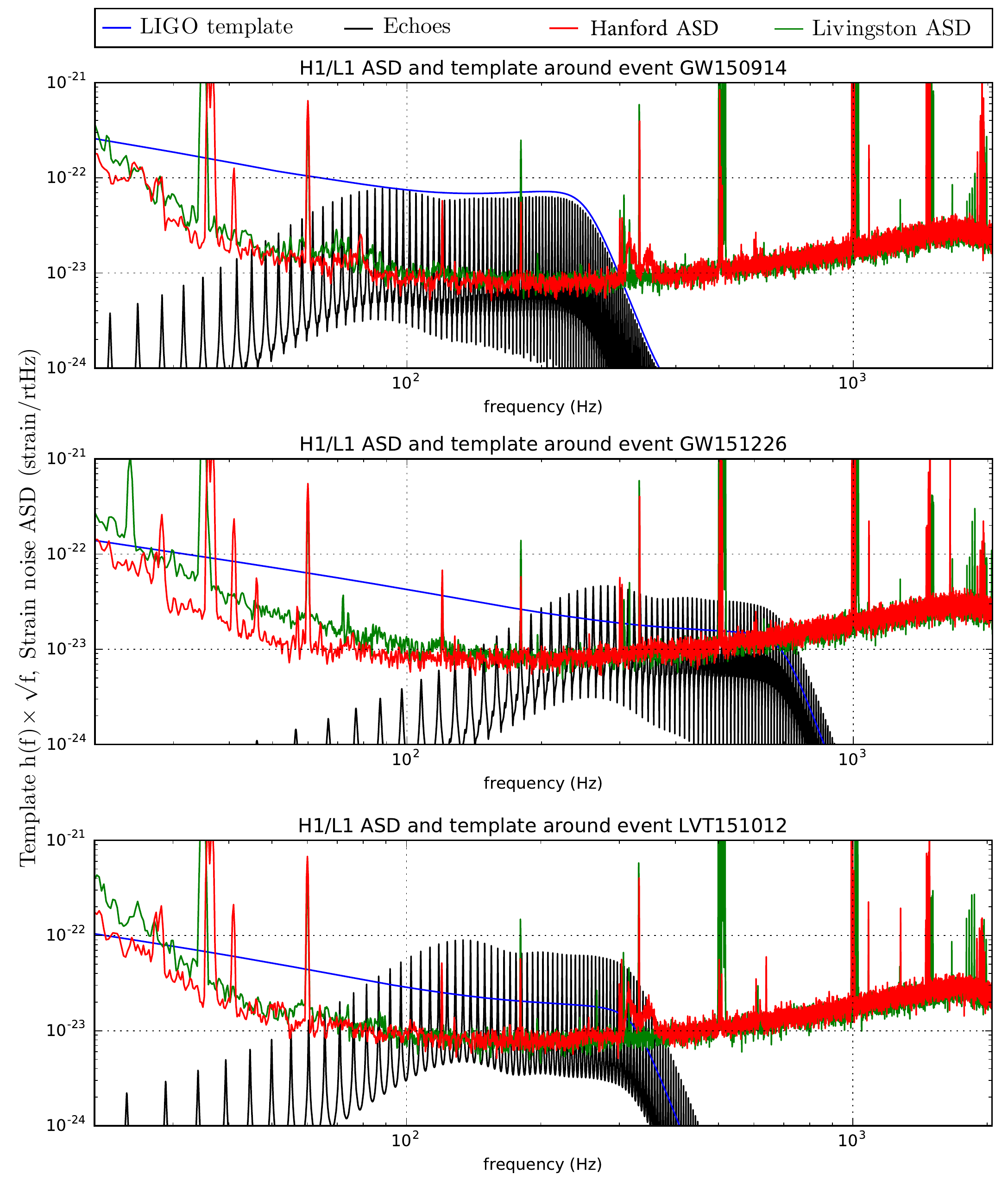}
    \caption{Best fit templates for LIGO main events and echoes (using the joint best fit described in the main text), in Fourier space (similar to Fig. \ref{template_matchfreq} in the main text). The amplitude spectral distribution (ASD) for each detector is shown for comparison.}
 \label{Match_frequency}
\end{figure}

\item Perhaps the most serious objection of Ashton, et al. concerns our estimation of significance, or false-detection probability (p-value). 
As we outlined in the introduction, it is already clear from Fig. (\ref{SNR_fig}) that the p-value for our SNR peak near $x=1$ should be $\lesssim 0.1$  and $\lesssim 0.01$, for GW150914 and combined events, respectively.

 The main criticism of Ashton, et al. stems from us quantifying our p-values (in the original arXiv submission of the main text) by considering how often random intervals of size $\Delta x =0.0054$ have an SNR bigger than the peaks we observe at $x=1.0054$, while we should have actually allowed for different choices of $\Delta x$. This would depend on the prior for $\Delta x$: the larger the prior, the higher would be the p-value.
 
 \begin{figure}[t]
  \centering
    \includegraphics[width=0.4\textwidth]{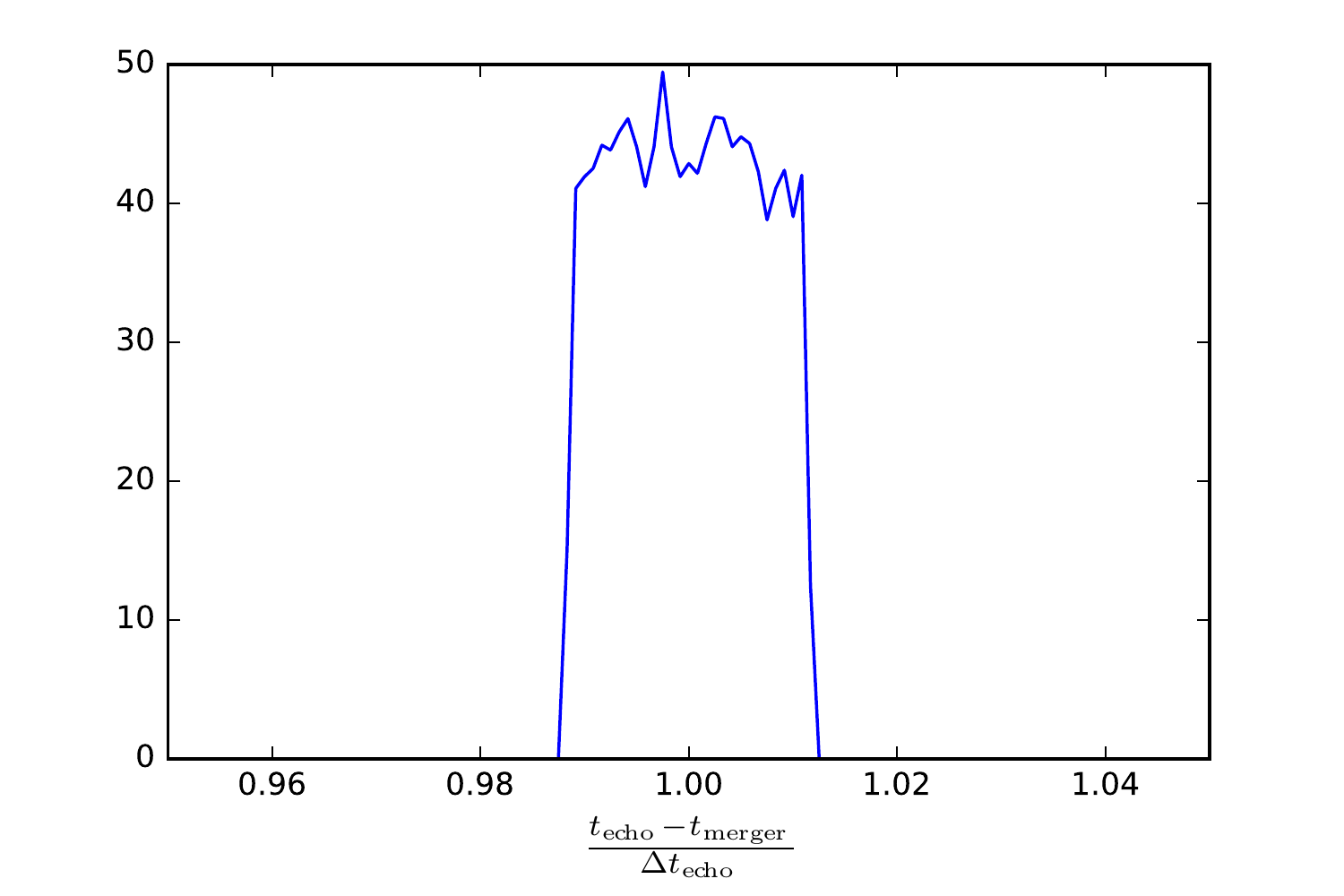}
    \caption{Resulting prior distribution on  $x=\frac{t_{\rm echo} - t_{\rm merger}}{\Delta t_{\rm echo}}$, assuming a random phase for the echo template. }
 \label{echo_prior}
\end{figure}
 
 However, we already have a decent idea about this prior from Eq. (\ref{nonlinear}) in the main text which suggests $\Delta x = {\cal O}(0.01)$, not far from what we used. We can get a more concrete handle on this prior by assuming that the echo template acquires a random phase (with respect to the main event) due to nonlinear propagation effects. Figure (\ref{echo_prior}) shows the resulting prior on $\Delta x$, which we find by replacing the data in our SNR computation (for GW150914) by the echo template with a random phase, and finding the position of the peak.  This results in a near top-hat prior with $-0.01 <\Delta x <0.01$ (an interval of $0.02$ rather than $0.0054$), which slightly increases the p-value to 0.011 (or significance of 2.5$\sigma$). 
 
Yet another way to quantify the significance would be to define a ``loudness'' function which averages the maximum likelihood for the echoes with a gaussian prior $x =1 \pm \sigma_{\rm echo}$, i.e. :
\begin{equation}
L(x,\sigma_{\rm echo}) \equiv \int \exp\left[{\rm SNR}_{\rm{total}}^{2}(x') \over 2\right] \times \frac{\exp\left[-\frac{(x-x')^{2}}{2\sigma^2_{\rm echo}}\right]}{\sqrt{2\pi \sigma^2_{\rm echo}}} dx'.\label{prior}
\end{equation}

We again use the LIGO data stream within the range 9-38 $\times  \overline{\Delta t}_{\rm{echo},I}$ after the merger event, to quantify how often $L(x,\sigma_{\rm echo})$ exceeds $L(1,\sigma_{\rm echo})$, for a given $\sigma_{\rm echo}$. This plotted in Fig. (\ref{p_value}), and provides an alternative p-value (or probability of false detection). This is also minimized at $\sigma_{\rm echo} \simeq 0.5\%$, with p-value of $0.01$ (or significance of $2.6\sigma$).
\begin{figure}[t]
  \centering
    \includegraphics[width=0.4\textwidth]{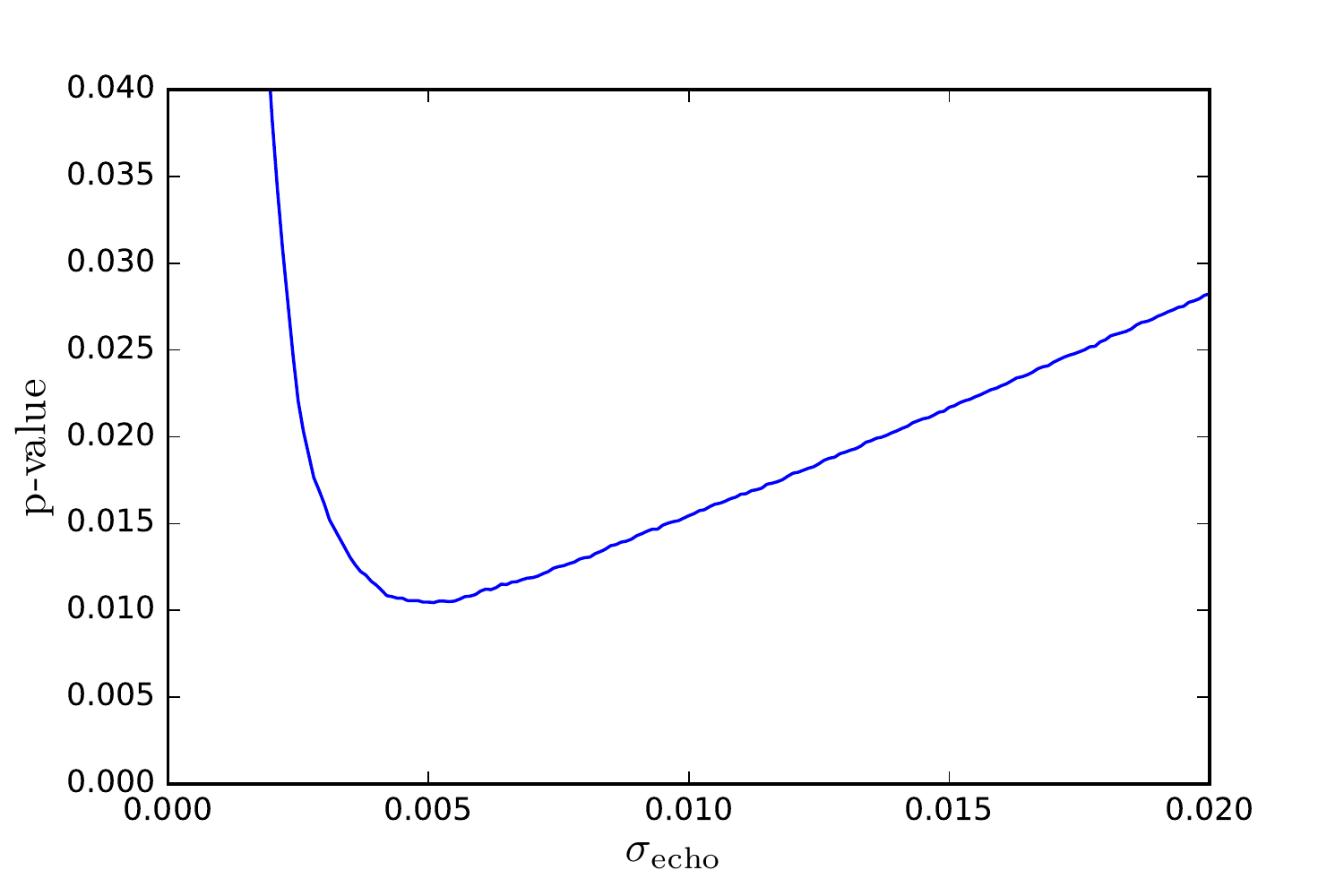}
    \caption{An alternative false detection probability (p-value) as a function of uncertainty in $t_{\rm echo}$ defined in Eq. (\ref{prior}). }
 \label{p_value}
\end{figure}

\item Ashton et al. are concerned that the range 9-38 $\times  \overline{\Delta t}_{\rm{echo},I}$ after the merger event, which we use to quantify false detection probability, might be contaminated by the echoes and somehow affect our significance estimation. Firstly, this is unlikely, as the evidence for echoes remains marginal and nearly all LIGO data (away from the merger event) is dominated by noise. Secondly, p-value quantifies the probability of null hypothesis, i.e. how often you see the echoes, assuming that there are none. As such, to find p-value one should assume that LIGO data, away from the main event, is pure noise and use that to quantify detection probability, which was what we did.  Therefore, we find this criticism ill-founded. 

Ashton et al. further advocate using larger stretches of LIGO data (which is publicly available) to define p-value more precisely. While this is in principle correct, LIGO noise is known to significantly vary and be very non-gaussian over long time-scales (see Fig's 14-15 in \cite{TheLIGOScientific:2016src}), which makes the interpretation of p-value ambiguous. The 9-38 $\times  \overline{\Delta t}_{\rm{echo},I}$ interval used is quite adequate to quantify the p-value for our signal, as otherwise we would see a sharp cut-off in our SNR cumulative distribution (Fig. \ref{Histogramloglog} in the main text). We tested this by looking at other stretches of data within a minute of the main events. As can be seen in Fig. \ref{Histogram-non-gaussian}, we observe that the spread in the p-values in the tail of the distribution is much higher than expected from Poisson statistics of the SNR peaks. Interestingly, the smallest p-value is obtained within the range closest to the main event. This might be expected as the marginal LVT151012 detection is preferentially close to a minimum of the LIGO (combined) detector noise. We believe this justifies using a smaller stretch of data close to the main events, to obtain a faithful reflection of the p-value using the ``instantaneous'' LIGO noise properties. 
\begin{figure}
  \centering
    \includegraphics[width=0.5\textwidth]{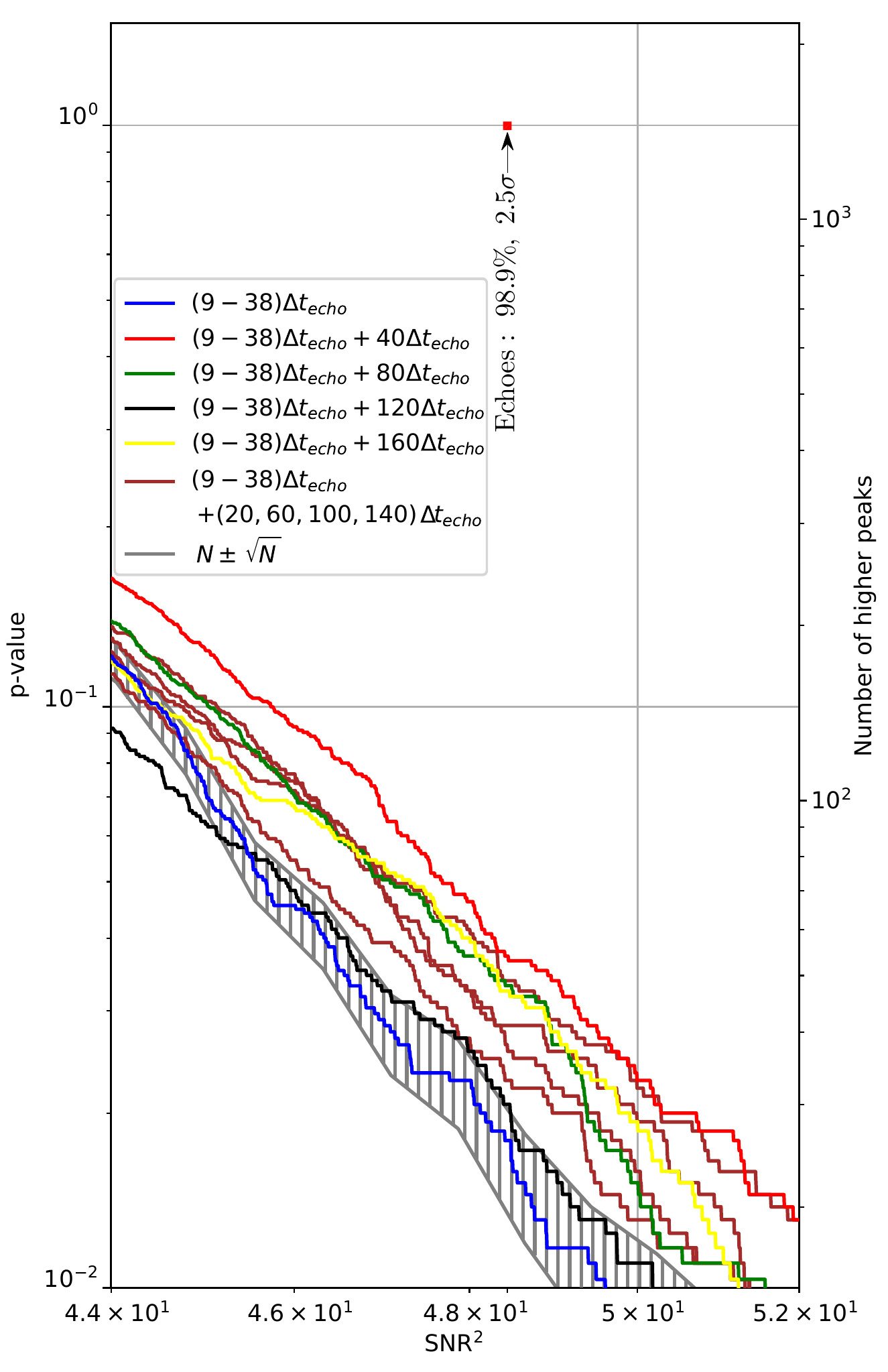}
    \caption{Distribution of p-values for combined events for different stretches of data within 1 minute of the main events. Surprisingly, the blue line which is closest to the main event, and is used to define p-value in the main text (Fig.  \ref{Histogramloglog}), happens to give the smallest p-value. The shaded region depicts the Poisson error range for blue histogram, showing that the variation in p-values is clearly much larger. We interpret this as non-gaussianity and/or non-stationarity in the LIGO noise properties. Here the y-axis on the left (right) shows p-value (number of higher peaks) within the mentioned range of data. The total number of ``peaks'' considered in each histogram is $(38-9)/0.02= 1450$. }
 \label{Histogram-non-gaussian}
\end{figure}

\end{enumerate}


\section{\label{The statistics of gaussian credible region}LIGO predictions for Echo time delays}

We approximate the uncertainty in the final redshifted mass $M$ and angular momentum $a$ of the LIGO black hole merger events by a gaussian probability distribution:
\begin{eqnarray}
\!\!\!\!\!\!\!\!P(\Delta a, \Delta M) \ \ \ \ \ \ \ \ \ \ \ \ \ \ \ \ \ \ \ \ \ \ \ \ \ \ \ \ \ \ \ \ \ \ \ \ \ \ \ \ \ \ \ \ \nonumber \\
\!\!\!\!\!\!\!\! =\frac{\sqrt{\det M}}{2 \pi} \exp(-\frac{1}{2} \alpha \Delta a^{2} - \frac{1}{2} \beta \Delta M^{2} -  \gamma \Delta a \Delta M)
\end{eqnarray}
where we assume $\langle M \rangle$ and $\langle a \rangle$ are the best fit reported values, while their inverse covariance matrix is given by,
\begin{eqnarray}
M_{ij}=
\begin{bmatrix}
    \alpha & \gamma\\
    \gamma & \beta
  \end{bmatrix}.
\end{eqnarray}
Here
\begin{eqnarray}
\Delta a = a - \langle a \rangle \ \ \ \ , \ \ \ \ \Delta M = M - \langle M \rangle.
\end{eqnarray}

We can then obtain the probability distribution of $\Delta t_{\rm echo}$,
\begin{eqnarray}
\!\!\!\!\!\!\!\!\!\!\!\!\!\!\!\!\!\!\!\!\! P(\Delta t) \ \ \ \ \ \ \ \ \ \ \ \ \ \ \ \ \ \ \ \ \ \ \ \ \ \ \ \ \ \ \ \ \ \ \ \ \ \ \ \ \ \ \ \ \ \ \ \ \ \ \ \ \ \ \ \ \ \ \ \ \nonumber \\
\!\!\!\!\!\!\!\! = \int \delta_{D}(\Delta t(a,M) - \Delta t_{\rm echo}) P(\Delta a, \Delta M) da dM \nonumber \\
=\int \frac{dM}{|\frac{\partial \Delta t}{\partial a}|}P(\Delta a, \Delta M)\ \ \ \ \ \ \ \ \ \ \ \ \ \ \ \ \ \ 
\end{eqnarray}
This leads to,
\begin{eqnarray}
P(\Delta t) \simeq 
\frac{\sqrt{\alpha \beta - \gamma^{2}}}{\sqrt{2 \pi (\alpha \mu^{2} + \beta + 2 \gamma \mu)} \left| \partial \Delta t / \partial a \right|_{\bar{a}, \bar{M}}} \nonumber \\
\!\!\!\!\!\!\!\!\!\!\!\!\!\!\!\!\!\!\!\!\! \times \exp(-\frac{1}{2} \frac{\alpha \beta - \gamma^{2}}{\alpha \mu^{2} + \beta + 2 \gamma \mu} \frac{(\Delta t - \Delta \bar{t})^{2}}{(\left| \partial \Delta t / \partial a \right|_{\bar{a}, \bar{M}})^{2}})
\end{eqnarray}
where $\mu=-\frac{\left. \partial \Delta t / \partial M \right|_{\bar{a},\bar{M}}}{\left. \partial \Delta t / \partial a \right|_{\bar{a},\bar{M}}}$.

Using contour of $50\%$ credible regions reported in \cite{TheLIGOScientific:2016pea}, we can obtain the angles of the eigenvectors of the covariance matrix. This gives a relation between $\alpha$, $\beta$, and $\gamma$:
\begin{eqnarray}
\!\!\!\!\!\!\!\!\!\!\!\!\!\!\!\! \tan (2\theta_{I}) = \frac{2\gamma_{I}}{\alpha_{I}-\beta_{I}}=\left\{
 \begin{matrix}
  0.013848 & I=GW150914 \\
  0.0072280 & I=GW151226 \\
  -0.0038272 & I=LVT151012
 \end{matrix}
 \right.\nonumber \\ \label{angle}
\end{eqnarray}
For the mean of the distribution, using the detector frame (or redshifted) masses we obtain (see \cite{TheLIGOScientific:2016pea}
 Table IV),
\begin{eqnarray}
\!\!\!\!\!\!\!\!\!\!\!\!\!\!\!\! \left( \bar{M}_{I}/M_{\odot},\ \bar{a}_{I},\ \Delta t_{pred}(\bar{a}_{I}, \bar{M}_{I}) \right) \nonumber \ \ \ \ \ \ \ \ \ \ \ \ \ \ \ \ \ \ \ \ \\
=\left\{
 \begin{matrix}
 67.8 & 0.68 & 0.29559s & I=GW150914 \\
 22.6 & 0.74 & 0.10246s & I=GW151226 \\
 42 & 0.66 & 0.17962s & I=LVT151012
 \end{matrix}
 \right.,\nonumber \\
\end{eqnarray}
while $\langle \Delta a^{2}\rangle$, and $\langle \Delta M^{2}\rangle$ are $68\%$ credible region (see \cite{TheLIGOScientific:2016pea}
 Table IV),
\begin{eqnarray}
\!\!\!\!\!\!\!\!\!\!\!\!\!\!\!\!\!\! \left( \langle \Delta M_{I}^{2}\rangle / M_{\odot}^{2},\ \langle \Delta a_{I}^{2}\rangle \right)=
\left( \frac{\alpha_{I}}{\alpha_{I} \beta_{I} - \gamma_{I}^{2}},\ \frac{\beta_{I}}{\alpha_{I} \beta_{I} - \gamma_{I}^{2}} \right) \nonumber \ \ \ \ \ \ \ \ \ \ \!\!\!\!\!\!\!\!\!\!\!\!\!\!\!\!\!\!\!\!\!\\
\!\!\!\!\!\!\!\!\!\!\!\! =\left\{
 \begin{matrix}
 4.6866 & 0.0012058 & I=GW150914 \\
 6.7632 & 0.0016633 & I=GW151226 \\
 36.091 & 0.0034056 & I=LVT151012
 \end{matrix}
 \right.\nonumber \\
\end{eqnarray}
These can be combined with $\theta_I$'s (\ref{angle}) to give:
\begin{eqnarray}
\!\!\!\!\!\!\!\! \left( \alpha_{I},\ \beta_{I}/M_{\odot}^{2},\ \gamma_{I}/M_{\odot} \right) \nonumber \ \ \ \ \ \ \ \ \ \ \ \ \ \ \ \ \ \ \ \ \ \ \ \ \ \ \ \ \ \ \ \ \ \ \ \ \ \nonumber \\
\!\!\!\!\!\!\!\!\!\!\!\!\!\!\!\!=\left\{
 \begin{matrix}
 1019.1 & 0.26221 & 7.0546 & I=GW150914 \\
 634.92 & 0.15615 & 2.2940 & I=GW151226 \\
 305.49 & 0.028826 & -0.58452 & I=LVT151012
 \end{matrix}
 \right.\nonumber \\
\end{eqnarray}
With these values we can obtain the gaussian posterior for $\Delta t_{\rm echo}$'s,
\begin{widetext}
\begin{eqnarray}
P(\Delta t_{I})
=\left\{
 \begin{matrix}
 \frac{77.187{s^{-1}}}{\sqrt{\pi}} \exp({-5957.8{s^{-2}} (\Delta t - 0.2925{s})^{2}}) & I=GW150914 \\
 \frac{61.372{s^{-1}}}{\sqrt{\pi}} \exp({-3766.5{s^{-2}} (\Delta t - 0.1013{s})^{2}}) & I=GW151226 \\
 \frac{25.357{s^{-1}}}{\sqrt{\pi}} \exp({-643.00{s^{-2}} (\Delta t - 0.1778{s})^{2}}) & I=LVT151012
 \end{matrix}
 \right.
\end{eqnarray}
\end{widetext}

\begin{figure}
  \centering
    \includegraphics[width=0.45\textwidth]{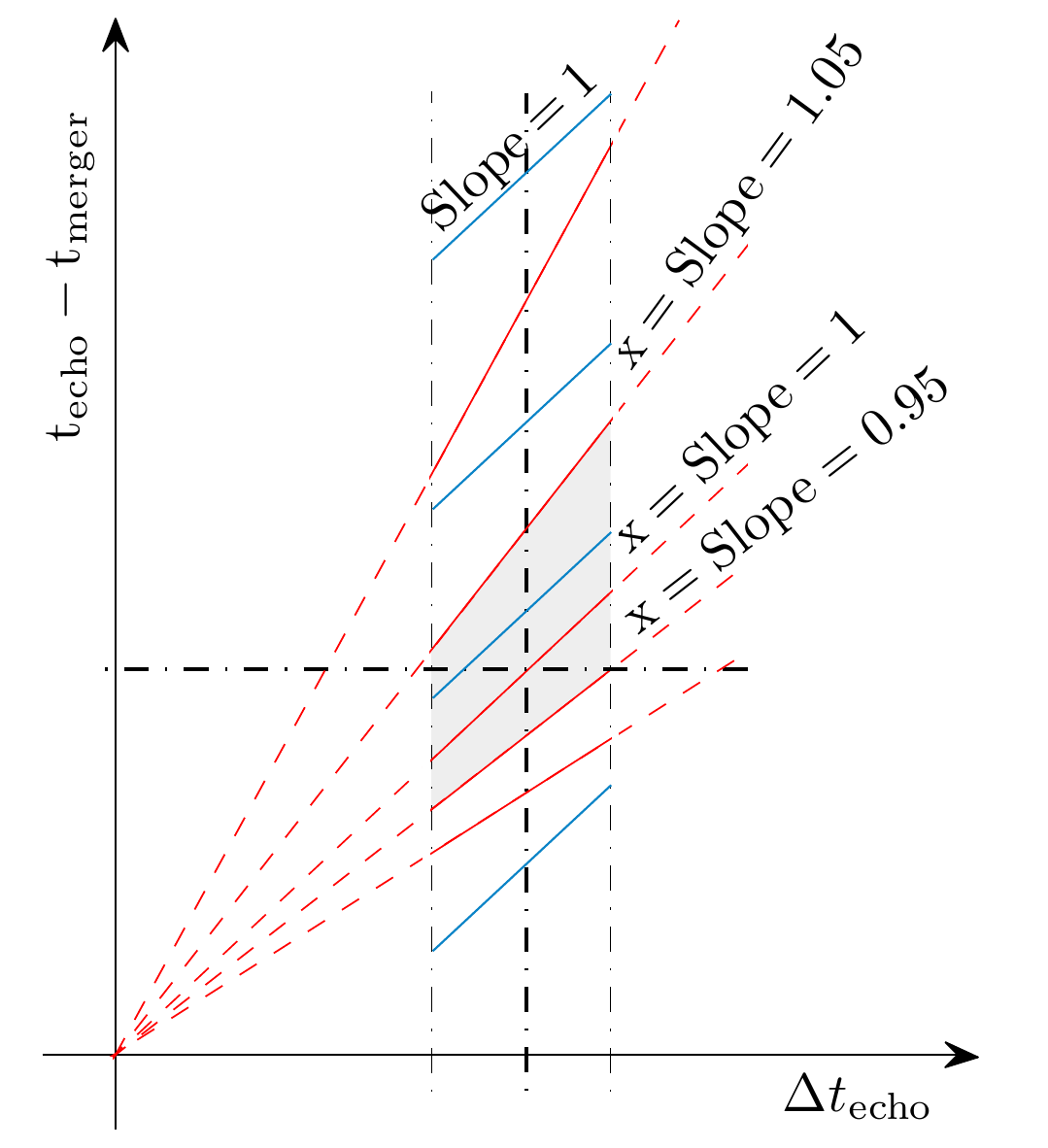}
 \caption{SNR peaks and their significance: We maximize SNR along the red lines for fixed values  of $x=\frac{t_{\rm echo} - t_{\rm merger}}{\Delta t_{\rm echo}}$, which is used to search for best-fit echo parameters, within $1\sigma$ region for $\Delta t_{\rm echo}$ (vertical lines). In particular, the grey trapezoid ($x=1 \pm 0.05$) corresponds to Fig. \ref{SNR} in the main text. After we find an SNR peak at some value of $x \simeq 1$, we quantify its significance by how often a comparable SNR$_{\rm max}$ can be found along the blue lines with unit slope for $x \gg 1$.}\label{Scale_Shift}
\end{figure}

\section{Discussion and Consistency of Statistical Methodology}

In this section we provide additional technical details that can be used to reproduce our results and test  their robustness and consistency. 

Figure (\ref{Scale_Shift}) demonstrates how we find our SNR peaks and their significance. In searching for echoes, we consider $x=(t_{\rm echo}-t_{\rm merger})/\Delta t_{\rm echo}= constant$ (red lines), and maximize SNR's varying all the free parameters within their priors (Tables I-II in main text). Fig. \ref{SNR} in the main text shows maximum SNR in the range $x=(0.95,1.05)$ (grey trapezoid in Fig.  \ref{Scale_Shift} ). Here $x$ depends on two variables:  $t_{\rm echo}$ and  $\Delta t_{\rm echo}$. As we discuss in the text, we seek the maximum SNR within $x = 1 \pm 0.01$ (see below), and quantify its significance by how often a higher SNR$_{\rm max}$ is achieved if the interval is shifted (e.g. the blue lines in Fig. \ref{Scale_Shift})  far from $x=1$.

We used 4 times higher ($4 \times 4096$ Hz = 16384 Hz) grid than the resolution of data for $\Delta t_{\rm{echo}}$ which is the most sensitive parameter in our search. For $t_{0}$ and $\gamma$ which are the less sensitive ones we used 76 and 100 points respectively.

Fig. \ref{echo_prior} shows how the best-fit SNR peak for the echoes of GW150914 moves if the echo template is multiplied by a random phase, expected from nonlinear effects during the merger.  We use this to fix the prior range for $t_{\rm echo}$, roughly corresponding to: 
\begin{equation}
x=\frac{t_{\rm echo}-t_{\rm merger}}{\Delta t_{\rm echo}} \simeq 1 \pm 0.01. \label{nonlinear2}
\end{equation}

Fig. \ref{SNR_fig} is the most clear demonstration of the significance of our results. This is the same as Fig. \ref{SNR} in the main text, but over a larger range.
For both GW150914 (the most significant reported LIGO event) and combined data from all three events, there exists a peak at distance $0.54\%$ from x = 1, which is inside the vertical grey bar (Eq. \ref{nonlinear2} and  Fig. \ref{echo_prior}).
For GW150914, the false detection probability is $11\%$ or the significance is $1.6\sigma$, meaning that comparable SNR peaks (from random noise) can be found within $\Delta x \simeq 0.02/0.1 = 0.2$, as can be seen with other peaks at $x \simeq 0.91$ and $1.16$. For the combined events, the false detection probability is $1\%$ or the significance is $2.5\sigma$, i.e. comparable peaks can only be found within $\Delta x \simeq 0.02/0.011 = 1.8$. This is also consistent with Fig. \ref{SNR_fig} as no higher peak can be seen within $\Delta x = 1.2$ in the plot.

Yet another consistency test is presented in Fig. (\ref{Hanford_Livingston}b), where we show independent SNRs for Hanford and Livingston detectors for the best fit parameters of GW150914 echoes. We can clearly see that not only the SNR peaks for echoes in different detectors coincide (after accounting for the event time-delays), but also their ratio is consistent with SNR ratios for the main merger event reported by LIGO collaboration.

\bibliography{Echoes_7_PRL}


\end{document}